\documentclass[a4paper,11pt]{article}
\pdfoutput=1 
\usepackage{jinstpub}
  
\usepackage{subfigure}
\usepackage{lineno}

\usepackage{color}
\usepackage{graphicx}
\usepackage[space]{grffile}

\usepackage{verbatim}
\usepackage{amsmath}
 
\usepackage{amssymb}
  
\usepackage{url}
\usepackage{bbold}
\usepackage{slashed}
\usepackage{epstopdf}

\usepackage{siunitx} 
\usepackage{float}

\usepackage{multirow}
\usepackage{threeparttable}
\usepackage{paralist}
\usepackage[percent]{overpic}

\usepackage{color}
\definecolor{darkgreen}{rgb}{0,0.5,0}
\definecolor{violet}{rgb}{0.5,0.5,1}

\begin{document}
\title{Vertex-Finding and Reconstruction of Contained Two-track Neutrino Events in the MicroBooNE Detector}
\author[ii]{P.~Abratenko}
\author[o]{M.~Alrashed}
\author[n]{R.~An}
\author[d]{J.~Anthony}
\author[hh]{J.~Asaadi}
\author[s]{A.~Ashkenazi}
\author[ll]{S.~Balasubramanian}
\author[k]{B.~Baller}
\author[t]{C.~Barnes}
\author[x]{G.~Barr}
% ...

% ...
\affiliation[d]{University of Cambridge, Cambridge CB3 0HE, United Kingdom}
% ...
\affiliation[k]{Fermi National Accelerator Laboratory (FNAL), Batavia, IL 60510, USA}
% ...
\affiliation[n]{Illinois Institute of Technology (IIT), Chicago, IL 60616, USA}
\affiliation[o]{Kansas State University (KSU), Manhattan, KS, 66506, USA}
% ...
\affiliation[s]{Massachusetts Institute of Technology (MIT), Cambridge, MA, 02139, USA}
\affiliation[t]{University of Michigan, Ann Arbor, MI, 48109, USA}
% ...
\affiliation[x]{University of Oxford, Oxford OX1 3RH, United Kingdom}
% ...
\affiliation[hh]{University of Texas, Arlington, TX, 76019, USA}
\affiliation[ii]{Tufts University, Medford, MA, 02155, USA}
% ...
\affiliation[ll]{Wright Laboratory, Department of Physics, Yale University, New Haven, CT, 06520, USA}

  \emailAdd{microboone\_info@fnal.gov}
\abstract{
We describe algorithms developed to isolate and accurately reconstruct two-track events that are contained within the MicroBooNE detector.     This method is optimized to reconstruct two tracks of lengths longer than \SI{5}{\centi\meter}.
This code has applications to searches for neutrino oscillations and measurements of cross sections using quasi-elastic-like charged current events. The algorithms we discuss will be applicable to all detectors running in Fermilab's Short Baseline Neutrino program (SBN), and to any future liquid argon time projection chamber (LArTPC) experiment with beam energies $\sim \SI{1}{GeV}$.  The algorithms are publicly available on a GITHUB repository\cite{githubcode}. This reconstruction offers a complementary and independent alternative to the Pandora reconstruction package currently in use in LArTPC experiments, and provides similar reconstruction performance for two-track events.}

\maketitle


\section{Introduction}
\label{intoPaper}
The MicroBooNE experiment is  currently taking data in the Booster Neutrino Beam (BNB) at Fermilab since October 2015, consists of a LArTPC located 450 m downstream of the proton beam target \cite{TDR}. The detector cryostat has a total capacity of 170 metric tons of liquid argon. The system comprises three major sub-detectors:  a time projection chamber (TPC) of (2.6$\times$2.3$\times$10.4) \si{\cubic \meter} for tracking, a light collection system for triggering and reconstruction of the precise interaction time, and a cosmic ray tagger (CRT) for improved cosmic background reduction. The detector has been described in detail in Ref. \cite{uBooneDetectorPaper} and \cite{uBooNECRTpaper}.

Neutrino events that are contained within the detector, and that are consistent with the final state signature of one muon and one proton, are of interest to a number of physics studies in MicroBooNE.   
Such a topology includes a large predicted  charged current quasi-elastic (CCQE) component and is important for the investigation of the MiniBooNE low energy excess \cite{MBLEE_2018} in MicroBooNE.
While the signal for the latter consists mainly of an electron and at least one proton, the normalization sample consists of a muon and at least one proton. In this case, the reconstruction is identifying candidate events that may be selected as  ``1$\mu$1$p$'' (one muon and one proton) events after particle ID for use in constraining nuclear effects in argon for $1e1p$ events.    
To obtain a large sample of contained 1$\mu$1$p$ events, the MicroBooNE collaboration has developed a reconstruction package specialized for contained two-track events.     The purpose of this package is to identify and reconstruct those events with two tracks, each longer than a specified length, emanating from a vertex; any number of shorter tracks may be attached to the same vertex.

The physics analyses that will make use of this code package employ both the TPC and light collection subsystems.  However the three-dimensional reconstruction code described here uses only the TPC information, and so we describe only this subsystem. 

\begin{figure}[t]
\center
\includegraphics[width = 0.8\textwidth]{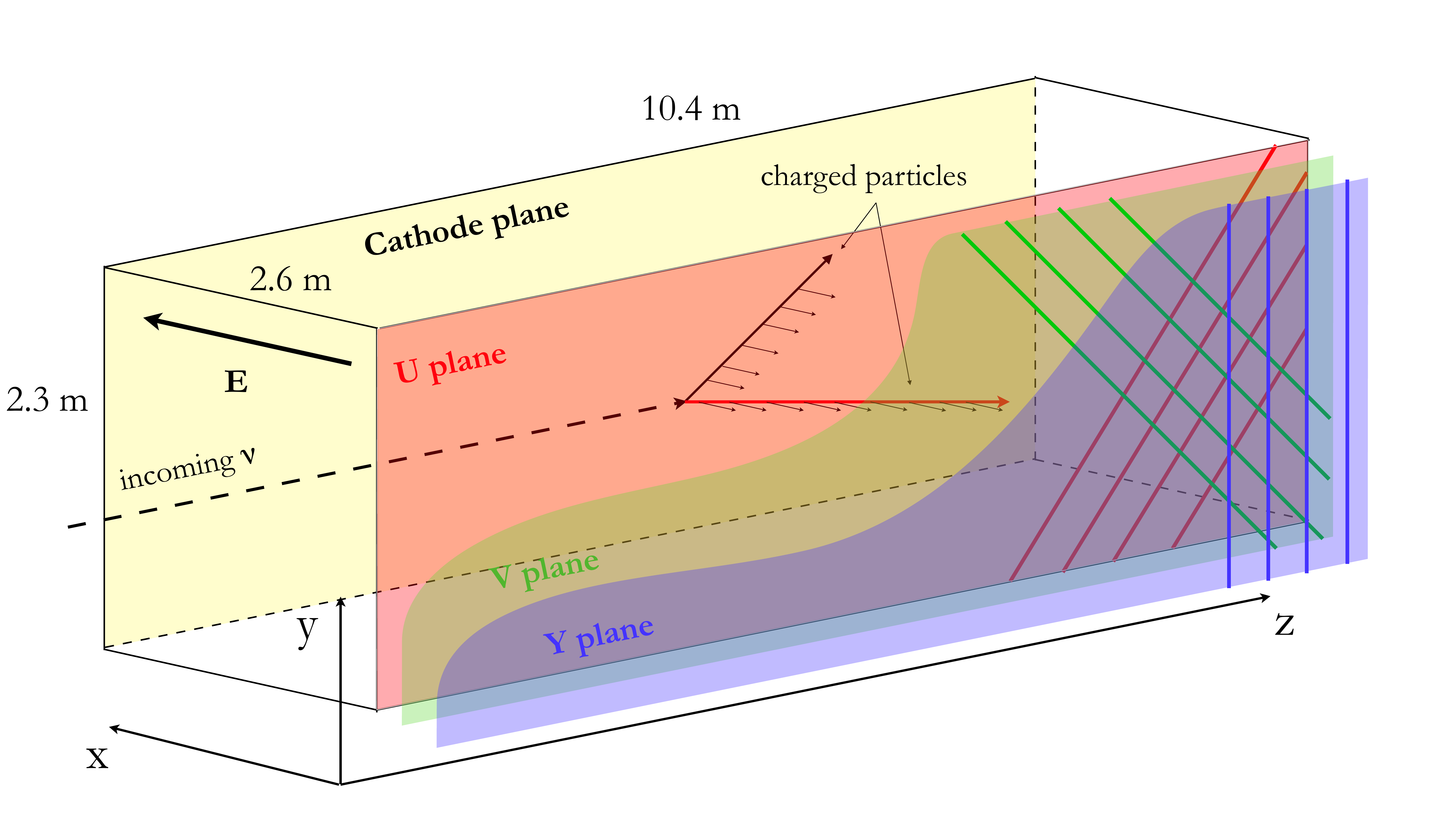}
\caption{\label{DetectorScematics} A schematic of the MicroBooNE TPC system. A volume of liquid argon has an applied electric field between a cathode plane and an anode plane. The anode plane is composed of three vertical planes of sense wires. The wires are oriented at $\pm 60^{\circ}$ and $0^{\circ}$ with respect to the vertical inside the (Y,Z) plane. Electrons from the ionization of charged particles drift towards the wire planes and are read out on the sense wires.}
\end{figure}

Figure \ref{DetectorScematics} illustrates the working principle of a LArTPC. Charged daughter particles from the neutrino-argon interactions ionize the argon along their path. Electrons from the ionization tracks drift in a horizontal \SI[per-mode=symbol]{273}{\volt \per \centi \meter} electric field along the $x$ axis, towards the vertical ($y$,$z$) anode plane composed of three parallel sense wire planes that provide the charge readout.

The signals from the three wire planes form three views, U (2400 wires at $+60^{\circ}$ from vertical),  V (2400 wires, $-60^{\circ}$), and Y (3456 wires, $0^{\circ}$). The U and V planes detect signals via electomagnetic induction, while the Y plane collects the charge of the ionization tracks. The wire spacing in each plane is \SI{0.3}{\centi \meter}. The wire waveforms are read out with a sampling time of \SI{0.5}{\micro \second}, and with a shaping time of the digitizers of \SI{2}{\micro \second}. This results in highly detailed event information that we exploit by treating the time versus wire waveform event display plots from each of the three planes as images with pixels, as described below. This use of high-resolution images allows the analysis chain to make use of deep learning algorithms. 
Indeed, in this article, we present an algorithm-based reconstruction approach leveraging the output of a pre-processing performed using a deep neural network.

\begin{figure*}[t]
\center
\includegraphics[width=0.8\textwidth]{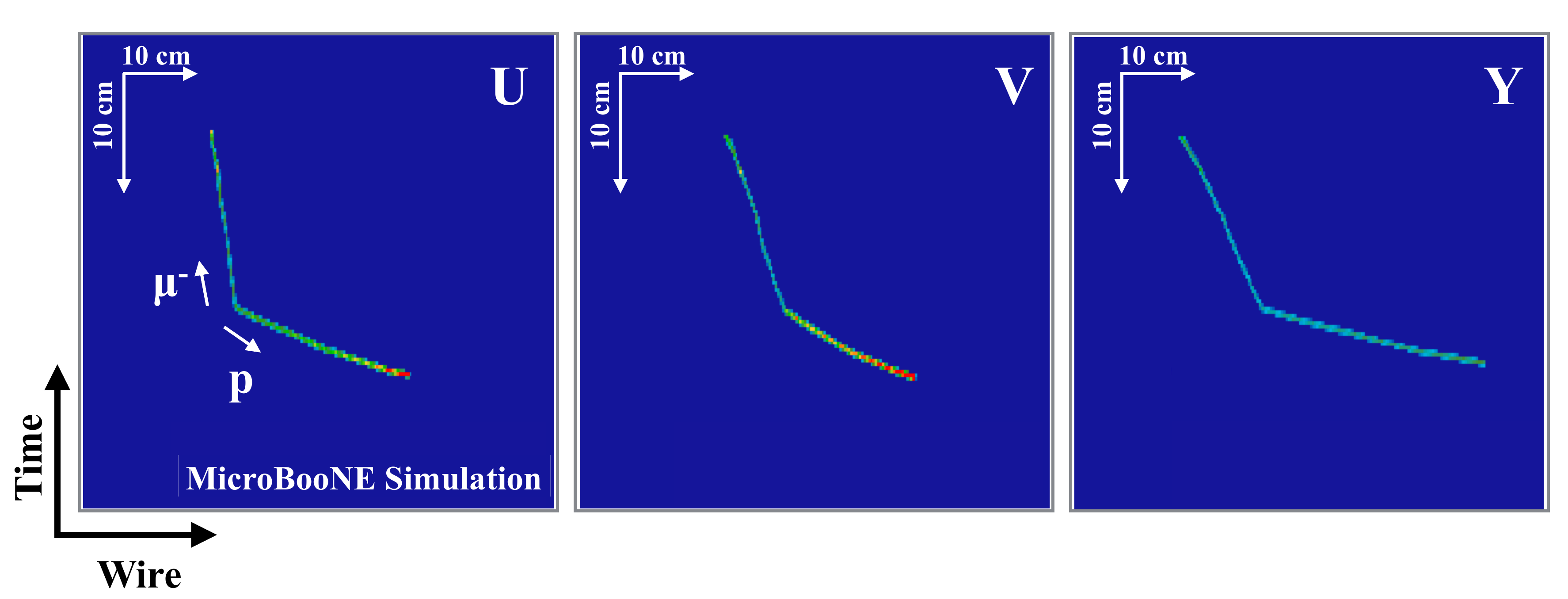}
\caption{
\label{CCQEexample} A simulated $\nu_\mu$ event shown in the three wire planes that illustrates the features of interest for this reconstruction package. The $\nu_\mu$ true energy is \SI{483}{\mega \electronvolt}, producing a single proton (deposited energy \SI{266}{\mega \electronvolt}) and muon (deposited energy \SI{73}{\mega \electronvolt}). The representation of the charge induced or deposited in each wire and per time sample creates an image of the interaction.}
\end{figure*}

Our reconstruction package is discussed within the context of MicroBooNE physics analyses.   
However, the approaches are generic to LArTPC detectors that run in a neutrino beam with an energy of the order of a \si{\giga \electronvolt}.  Examples of such experiments are SBND and ICARUS, which will run in the same BNB neutrino line as MicroBooNE in the near future \cite{SBN}.  As an example, SBND is expecting  $\sim$3~M $\nu_\mu$ CCQE events for an exposure of $6.6 \times 10^{20}$ protons on target (POT), of which $\sim$2~M with the final state $1\mu 1p$ \cite{SBN}. The approach is also appropriate for reconstructing atmospheric neutrino events in the DUNE near and far detector, as well as lower energy beam-based events \cite{DUNE}, although in DUNE far detector, complications due to cosmic rays will be substantially reduced compared to the surface-based detectors on the BNB. About 18 M $\nu_\mu$ CCQE events are expected per year in DUNE liquid argon near detector, with $\sim$13.7 M/year events with the $1\mu 1p$ final state\cite{DUNENDprivate}.

We present results based on simulated particles at energies relevant to the MicroBooNE beam using MicroBooNE's Monte Carlo simulation package (MC).
MicroBooNE uses GENIE \cite{GENIE} to simulate the initial neutrino interactions in argon. For the purpose of this article, we only make use of predicted $\nu_\mu$ CCQE interactions with their final states containing a muon of at least \SI{35}{\mega\electronvolt} and only one proton of at least \SI{50}{\mega\electronvolt}; any number of particles below these thresholds can be present, as well as non-ionizing particles such as neutrons and $\gamma$.
The outgoing particles are fed to a GEANT4 simulation  \cite{Geant4_1, Geant4_2,Geant4_3} of the detector.
Data from unbiased off-beam readouts, containing only comic rays, are overlaid on top of the simulated event displays containing the tracks from the neutrino interactions. This ensures a correct representation of the cosmic ray background in our simulated neutrino events.

\section{Data Pre-processing}

An essential and very difficult problem to solve in reconstruction of LArTPC data on the surface is identification and removal of cosmic rays  from the events.   With minimal overburden, MicroBooNE averages 12 cosmic rays in a 2.3 millisecond readout window. The high rate of cosmic rays makes this background a complication for neutrino analyses. 
To address this issue,  prior to three-dimensional reconstruction, an algorithm is applied to tag pixels corresponding to cosmic rays.    This code will be described in a future separate article, and so is only briefly described here. Cosmic rays are identified by their boundary-crossings at the edges of the active region. Through-going cosmic rays cross two boundaries. The cosmic ray tagging algorithm starts at the boundary and works inward, into the detector, labeling consecutive charge depositions, finding a path between two boundary crossings.
Once all charge that is connected to a boundary is identified, a 3D volume in the TPC is found for which the projection on each plane encompasses the remaining untagged charge clusters. Such a volume is called ``contained regions of interest'' (cROIs). These cROIs are then passed to the three-dimensional reconstruction code. Typically about 10 cROIs are found per event.  It is possible for tagged cosmic ray charges to appear within a cROI.

The cROIs are then passed to a deep-learning algorithm called a semantic segmentation network (SSNet) \cite{SS_1,SS_2}. The uses and performances of semantic segmentation in MicroBooNE analyses are described in detail in \cite{SSnetpaper}. For the purpose of this work, it should suffice to say that the Semantic Segmentation labels all pixels in the cROI image as either background, track-like, or shower-like, where background refers to empty pixels with little to no charge deposition. Typically, the semantic segmentation network will classify muons, charged-pion and protons as track-like; and electrons and photons as shower-like.

In summary, the inputs to the code we describe here are cROIs that have all charge tagged as cosmic-ray-, track-, shower-, or background-like. 
This information is used to find a three-dimensional vertex using the sets of images from the three planes. 
At that point a determination is made as to whether there are two, and only two, track-like chains of charge with lengths that pass our  requirements.
The pixels associated with the track-like chains are clustered. 
The  three-dimensional vertex is then passed to an algorithm that reconstructs the three-dimensional tracks. 
For the track-reconstruction stage, the pixel identification tagging is not used. 
We describe each of these steps below, and present information on the reconstruction efficiency and other figures of merit. 
First, we describe the input images that are utilized by this package.


\section{Using Images in the Reconstruction Package}

This reconstruction package treats MicroBooNE data and Monte Carlo event readouts as ``images''. 
By this, we mean that the TPC data from each of the three views are represented on a 2-dimensional plot, with wire number along the $x$ axis and drift time along the $y$ axis. 
The choice to analyze the detector in an image-format allows the use of widespread and very powerful computer vision tools such as ``OpenCV'' (open source computer vision) \cite{OpenCVpage}, which is a C/C++ based framework for computer vision that provides useful classes/functions for image processing.
It is a widely used application for pattern recognition.
Using images also allows for the implementation of deep learning algorithms at two points in the analysis. 
The first is the Semantic Segmentation, which precedes this reconstruction package, and was discussed above. 
The second is a convolutional-neural-network-based particle identification, which follows the reconstruction package presented here, and is beyond the scope of this article. 
The output of the three-dimensional reconstruction package described here can be used by deep-learning-base particle identification networks as presented in Ref. \cite{pidpaper}.
A likelihood-based particle identification that follows this package is also under development in MicroBooNE.

However, the value of using images goes beyond the applications of deep learning.  When constructing the simulated events, it is easy to overlay the pixels associated with a simulated neutrino event onto a real out-of-beam image or an image from the cosmic-ray simulation. 
Also, crucially for the three-dimensional reconstruction, information can be straightforwardly associated with each pixel at each stage of the algorithm, and then carried through to the following stages.   

The reconstruction package primarily uses two kinds of images. 
The first is the ``ADC image,'' which contains information on the charge in each pixel. 
The second is the ``SSNet image'' which contains information on whether pixels representing connected chains of charge are track-like, shower-like, or neither. 
In both cases, the cosmic-ray tagged pixels identified in the early stages are masked, and so are not visible in the images. 
Two additional pieces of information are also provided to the algorithm. 
The first are images that show the cosmic-ray-tagged pixels. 
The second are images indicating unresponsive wires.

\subsection{ADC images}

In the case of the ADC image, the intensity of each ``pixel'' is determined by summing the amplitude of the noise-filtered, deconvolved signal \cite{signalprocesspaper} from the wires over six time ticks.  This choice comes from the fact that, at 0.5 microseconds per tick, six time ticks is 3 microseconds, only slightly larger than the 2 microsecond shaping time of the ASICs. Also, at the current drift field, the nominal drift velocity of the electrons is $1.098$ \si[per-mode=symbol]{\milli \meter \per \micro \second} \cite{Adams:2019qrr}. Therefore, the drift distance over six time ticks is about \SI{0.33}{\centi \meter}, which is similar to the detector's \SI{0.3}{\centi \meter} wire pitch.   The images are created using the LArCV code available on GITHUB \cite{githubcode}. An image is created for each cROI in each of the wire-plane views.

Figure \ref{CCQEexample} shows an example of a simulated $\nu_\mu$ CCQE event, of interest for this reconstruction package.   An ADC image is made for each plane, as illustrated by each of the three image frames. The size of each image is set by the cROI-finding algorithm.  The neutrino in this event has a true  energy of $E_\nu^\text{true}=\SI{483}{\mega \electronvolt}$; the kinetic energies of the emitted muon and proton are \SI{73}{\mega \electronvolt} and \SI{266}{\mega \electronvolt}, respectively, which is typical of the kinematics this package was designed to reconstruct.

\subsection{SSNet images}

\begin{figure}[t!]
\center
\includegraphics[width=0.7\textwidth]{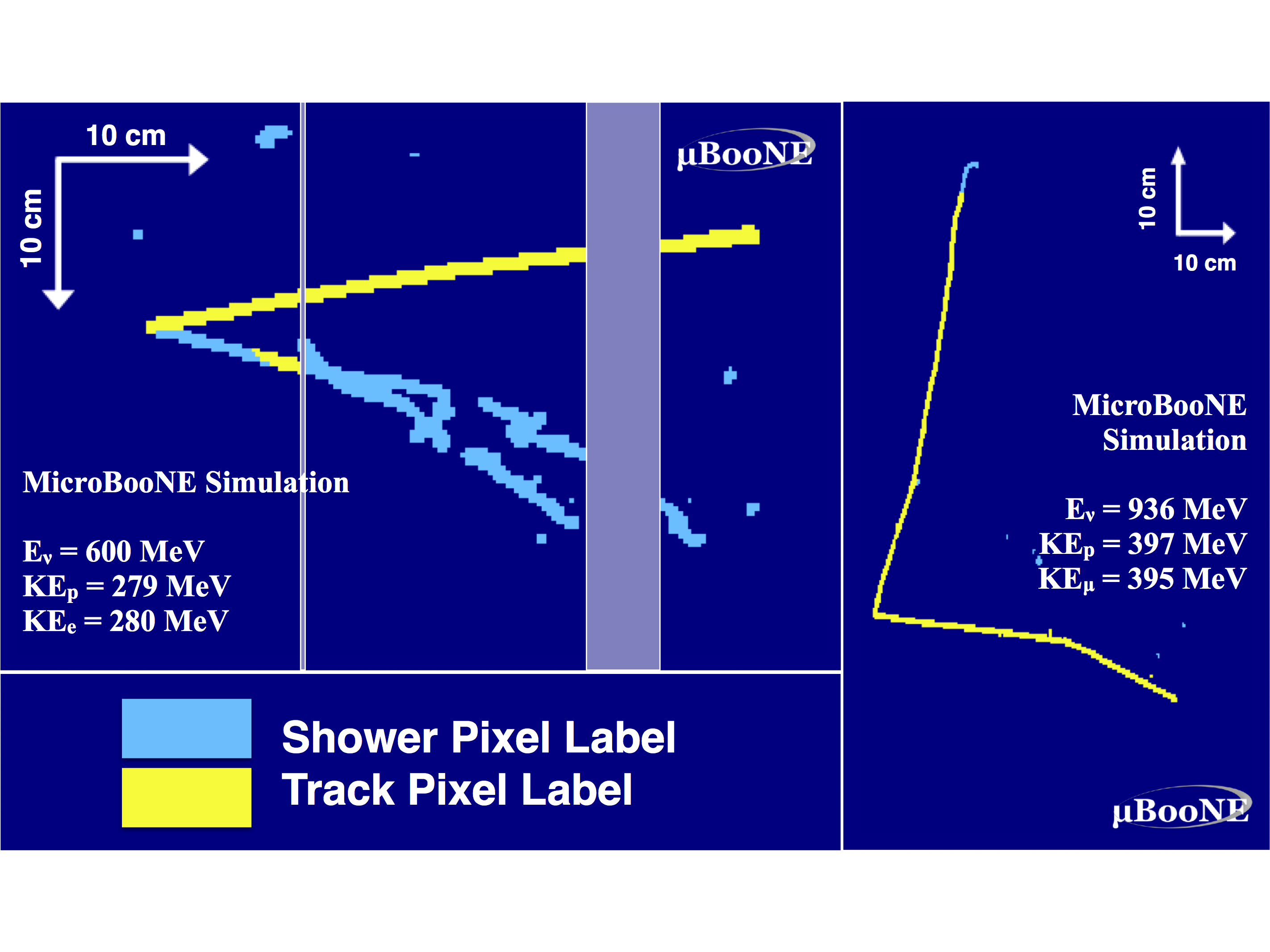}
\caption{
\label{OpenCVcorr} Two examples illustrating the SSNet image pixel labeling. The left panel shows a $1e1p$ type event from a $\nu_{e}$ interaction. The original neutrino energies as well as the kinetic energy of daughter particles are indicated on the figure. The pixels corresponding to the proton are  correctly classified as track-like by the SSNet. The pixels corresponding to the electron are mostly classified as shower-like by the SSNet, except for a small portion mistakenly labeled as a track. The right panel shows a $1\mu1p$ type event from a $\nu_\mu$ interaction. Both the proton and muon tracks are correctly labeled as track pixels. The muon decays into a Michel electron, classified as shower pixels. Dark blue pixels correspond to empty pixels, without charge deposition. Clear vertical bands correspond to simulated unresponsive regions of the TPC.}
\end{figure}

The SSNet images, which are also constructed for each cROI and for each wire plane view, are created by feeding the ADC images to a SSNet. The SSNet identifies the pixels based on their surroundings into three categories:
\begin{itemize}
\item track pixels
\item shower pixels
\item background pixels
\end{itemize}

Two examples of SSNet outputs are shown in \mbox{Fig. \ref{OpenCVcorr}}. The left panel shows a view of a $1e1p$ $\nu_{e}$ interaction in the Y plane, with a \SI{600}{\mega \electronvolt} neutrino producing a \SI{279}{\mega \electronvolt} proton and a \SI{280}{\mega \electronvolt} electron. The proton track is correctly classified as containing only track-like pixels (in yellow), and the electron shower is mostly classified as shower pixels (in light blue). A small fraction of the shower pixels are mistakenly labeled as track-like. Background pixels, corresponding to pixels without charge deposition, are shown in dark blue. 

 The right panel shows a Y plane view of the interaction of a \SI{936}{\mega\electronvolt} $\nu_\mu$  producing a $1\mu1p$ event with a \SI{397}{\mega \electronvolt} proton and a \SI{395}{\mega \electronvolt} muon, which decays into a Michel electron. The Michel electron is classified as shower pixels, while the proton and muon tracks are correctly labeled as track-like.\\

In this note, we will focus on the $1\mu1p$ topology; therefore we will be looking for two-track vertices, neglecting track-shower vertices.

\subsection{Cosmic-ray-tagged images}

The reconstruction algorithms described in the rest of this article are also supplied with an additional image with cosmic-ray information. The through-going muon pixels in this image are tagged, but remain visible to the algorithms. These pixels can be optionally removed from the ADC, track, and shower images to help reduce the probability that the algorithms will reconstruct a cosmic-ray background.

\subsection{Unresponsive-wire images}

In addition, the vertexing algorithm is supplied with an image marking the spatial locations of unresponsive wires, which can be time-dependent.  Providing this information allows the algorithm to know precisely which region in the image represents pixels which contain no charge, and make a decision about whether to veto this region, and potentially neighboring regions, for vertex finding and particle clustering.


\section{3D Vertex Finding and Particle Clustering \label{ThreeVx}}

This reconstruction step finds the 3D vertex and then clusters pixels belonging to individual particles.    In this algorithm, the pixels tagged as cosmic rays are removed from the images.      
For each of the three views, a set of three images are provided to this algorithm. The first image contains all pixels in the cROI (ADC image), the second contains pixels labeled as tracks (track image) after the SSNet correction, and the third image contains pixels labeled as showers (shower image).   

Along with the use of LArOpenCV, described in the previous section,  this code  makes use of a custom OpenCV package called Geo2D. This package has convenient tools for 2D geometrical analysis to supplement and extend OpenCV built-in data types.

\subsection{Vertex Finding using Track-identified Pixels \label{trackvx}}

This step makes use of only the track-identified pixels (the track image) to reconstruct a 3D vertex. 
The algorithm searches for a coincident ``vee" shape feature, as shown in  Fig. \ref{CCQEexample}, across the three wire planes which could indicate the presence of a $1\mu1p$ interaction. This event will be used throughout this section to illustrate the vertex reconstruction algorithm. The algorithm begins by identifying a collection of vertex ``seeds" in each plane. Vertex seeds are pixel locations in the image where a likely vertex may be present, for example at a kink point, the location where two tracks meet. The algorithm identifies vertex seeds by breaking down continuous sets of track clusters into smaller clusters which contain straight segments of charge.

\subsubsection{High-charge and low-charge clusters}

\begin{figure}[!t]
\center
\subfigure[]{\raisebox{1 pt}{\begin{overpic}[width = 0.32\textwidth]{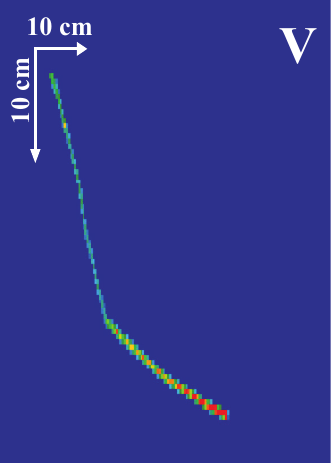}\put(1,102){\textbf{\scriptsize MicroBooNE Simulation}} \end{overpic}} }
\subfigure[]{\includegraphics[width = 0.32\textwidth]{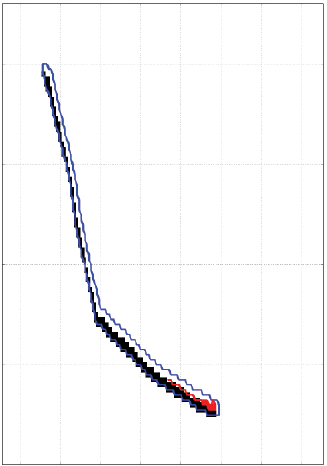}}
\caption{
\label{3d_vertex_hip_mip} (a) V-plane view of the $1\mu 1p$ event from Fig. \ref{CCQEexample}.  (b) Low charge (blue) and high charge (red) contours are found in this example $1\mu1p$ event. The high charge contour clusters a collection of pixels which have a high pixel intensity (here belonging to the proton Bragg peak). The low charge contour encloses all pixels in this track image as they are all above the 10 ADC threshold.}
\end{figure}

First, a threshold of 10 ADC is applied on the image to remove pixels from waveform deconvolution artifacts. A distinction between pixels in the low-charge (LC) and high-charge (HC) regimes as shown in Fig. \ref{3d_vertex_hip_mip} is then performed. The division between LC and HC pixel ADC count is a constant threshold per plane and is determined from a study of the pixel intensities of MC protons. The threshold value for the U, V, and Y planes are set at 140 ADC, 120 ADC, and 80 ADC, respectively. These values correspond to $10\%$ of the average pixel value for a proton track on each plane. Once the pixel ranges are separated, the algorithm finds groups of LC and HC pixels by applying the OpenCV contour finder. 

\subsubsection{Linear cluster breakdown}
Next the algorithm performs a shape analysis by breaking down LC and HC clusters, which are not linear, into linear clusters. For example, the blue LC contour shown in Fig. \ref{3d_vertex_hip_mip} has an obvious bend or ``kink" in it, which is associated with a possible neutrino interaction vertex. 
For each cluster, the algorithm computes the ``convex hull'' which is the smallest convex polygon that bounds the original cluster.\\

\begin{figure}[!t]
\center
\subfigure[\label{3d_vertex_hull_a}]{ \begin{overpic}[width=0.32\textwidth]{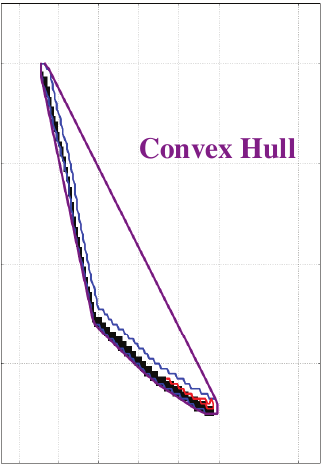} \put(1,102){\textbf{\scriptsize MicroBooNE Simulation}}\end{overpic} }
\subfigure[\label{3d_vertex_hull_c}]{\includegraphics[width=0.32\textwidth]{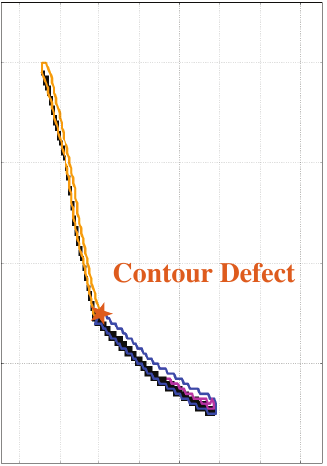}}
\caption{
\label{3d_vertex_hull} 
(a) The convex hull (purple) is computed for the LC contour (blue) of the event of Fig. \ref{3d_vertex_hip_mip}.  A contour defect is found on the LC contour, more than 5 pixels away from the corresponding convex hull edge. The contour defect indicates the location where the cluster is bending.
(b) The LC cluster is then broken into two clusters at the contour defect. The breaking procedure is carried out for each contour until no contour defects remain on the cluster. }
\end{figure}

Figure \ref{3d_vertex_hull_a} shows an example of convex hull (purple polygon) encompassing the pixels of a $1\mu 1p$ simulated event. The algorithm identifies the side of the convex hull that is furthest away from the cluster of pixels. 
The point on the contour that is furthest away from the corresponding hull side is called the ``contour defect", and is a location where the cluster is potentially bending and changing direction. If the convex hull side is far enough away (5 pixels) from the contour defect, the contour is then broken into two at the contour defect. Figure \ref{3d_vertex_hull_c} shows the three clusters obtained after this stage, 1 HC cluster (purple line) and 2 LC clusters (yellow and blue lines), as well as the contour defect of the original LC cluster (orange star). The algorithm then iteratively breaks down all clusters into linear segments as long as a contour defect can be found further than 5 pixels away from the convex hull.  The collection of contour defects are the first set of vertex seeds.\\

\begin{figure}[!t]
\center
 \begin{overpic}[width=0.32\textwidth]{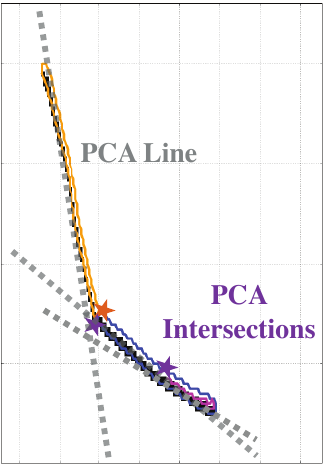} \put(1,102){\textbf{\scriptsize MicroBooNE Simulation}}\end{overpic}
\caption{
\label{pca_xing} Diagram showing the calculation of the PCA for each track cluster. For each broken cluster found in Fig. \ref{3d_vertex_hull}, a 2D PCA line is calculated. Three PCA lines (gray dashed) are found in this example, one per cluster. 
The points where the PCA lines cross are shown as purple stars and are called PCA intersections.  
The orange star is the vertex seed from the earlier contour defect stage.}
\end{figure}

The second set of vertex seeds is found using a principal component analysis (PCA) procedure which calculates the best straight line formed by the clusters of pixels. A PCA is calculated for each of the clusters found by the bounding box stage separately. Since all clusters have been broken into linear segments by the previous step, a linear approximation is applicable. The algorithm then computes the intersection of all possible PCA lines on the plane. If the lines intersect near a location with charge, the point is saved and is added to the set of vertex seeds. Intersection points located in pixels with no charge are ignored. Figure \ref{pca_xing} shows the three PCAs found in the example event. Although three intersection points are found, only two correspond to pixels with charge and are kept.
This type of vertex seed helps find the 2D location where tracks which may be changing direction. Also, using a linear approximation for the clusters gives an additional set of points beyond the defects points, increasing the efficiency of finding the actual vertex.\\

\subsubsection{Finding 3D-consistent vertex candidates}
The final set of vertex seeds is composed of both the contour defects and the PCA intersections. The proximity of the seeds from the two methods is a strong indication of the actual vertex location. 
Each of these 2D points is considered a vertex ``seed" and serves as a starting point for 3D vertex search.\\

The algorithm makes use of the fact that a correct vertex will appear near the same time tick in each view. This reduces the seed sample to the time-coincident ones. The X position of these candidates can be determined by using the trigger time and the known drift speed to match the time tick to an X position. The Y and Z positions can then be determined by using wire coincidences between two or three planes.

The algorithm then performs an exhaustive search for a 3D vertex around each seed by minimizing an angular metric. 
This angular metric is a single quantity which evaluates the likelihood of two tracks being emitted radially outward at the same position across two or more planes for a given vertex seed. This metric is minimized when a point in 3D space is found where the 2D projections indicate that the particles are coming out of a single point.

\begin{figure}[!t]
\center
\subfigure[\label{3d_zoom_theta}]{\begin{overpic}[trim = 0 0 18cm 0, clip, width=0.15\textwidth]{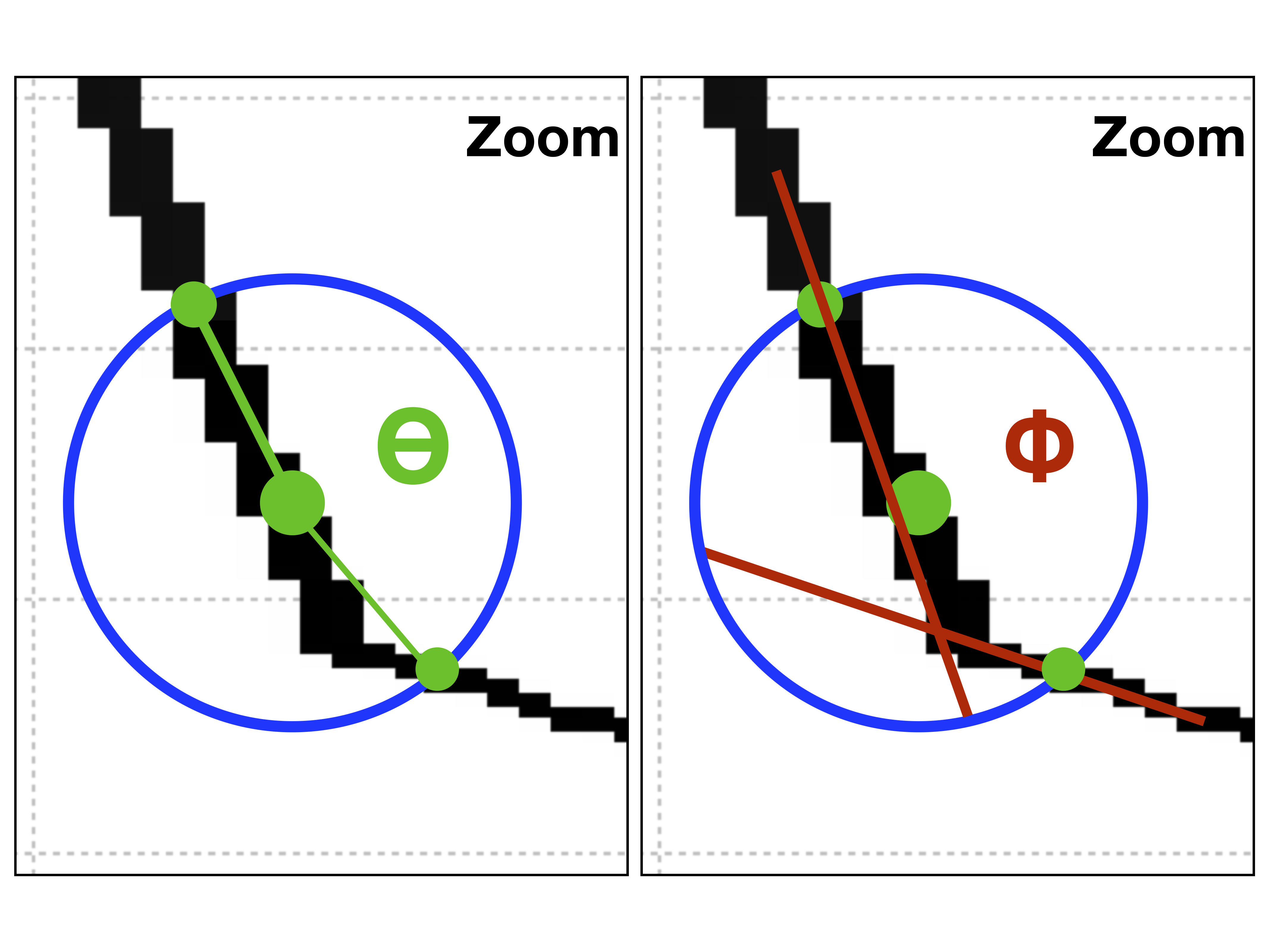}\put (2,96){\textbf{\scriptsize MicroBooNE Simulation}}\end{overpic}}
\subfigure[\label{3d_zoom_phi}]{\includegraphics[trim = 18cm 0 0 2cm, clip, width=0.15\textwidth]{vtx_algo/06_3d_zoom_3.png}}
\subfigure[\label{3d_zoom_large}]{\begin{overpic}[trim = 0  0  0  0, clip, width=0.43\textwidth]{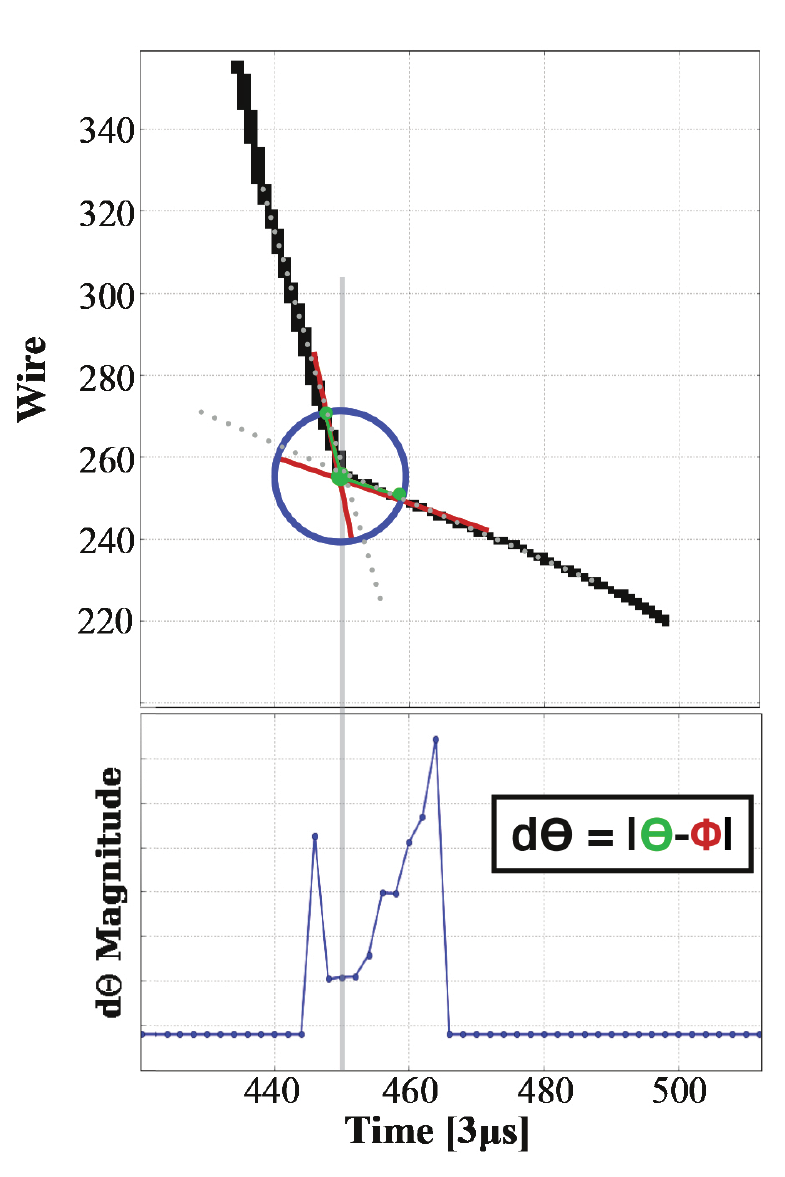}\put (11,101){\textbf{\scriptsize MicroBooNE Simulation}}\end{overpic}}
\caption{
\label{3d_zoom}
Evaluation of the angular metric used to find the best seed position. For each candidate vertex position found in \mbox{Fig. \ref{pca_xing}}, a circle is drawn centered at the vertex candidate position, the tracks intersects the circle boundary at two points. 
(a) and (b) describe the angles of interest $\theta$ and $\phi$. 
The circle center is stepped in increments of one pixel along the initial straight line PCAs in (c) and the magnitude difference, $d\Theta$, between $\theta$ and $\phi$ is recorded per six-time-tick bin (3 microseconds). 
The bottom plot shows the resulting spectrum for the scanning procedure applied centered at the vertex seed near the kink point. The graph is $d\theta$ versus time. 
}
\end{figure}

The algorithm begins a search for a 3D vertex by using images from all three wire planes. The following operation is performed for every vertex seed on each plane. 
Circles of radius of 6 and 12 pixels are drawn with the given vertex seed at the center. The circle size with the greatest number of clusters intersecting it is used. In cases where the same number of clusters cross the circles, the largest circle is kept. The algorithm then identifies the points at which the out-going clusters cross the circle. 
For each pair of out-going clusters, the algorithm evaluates two angular quantities:  
\begin{itemize}
\item The first angular quantity, $\theta$, is the smallest angle between the center of the circle, and each pair of cluster-circle intersection points.
\item The second quantity, $\phi$, is calculated in the following way. At each cluster-circle intersection point a small region of 7$\times$7 pixels is identified.  In this region, a PCA of charge carrying pixels is computed, and provides an estimate of the local cluster directions. $\phi$ is the smallest angle between each local PCA pair.
\end{itemize}
The definitions of the two angles are illustrated in Fig. \ref{3d_zoom_theta} and Fig. \ref{3d_zoom_phi} respectively.

Starting at the vertex seed, two line segments are drawn from the initial local PCA approximations, as illustrated in Fig. \ref{3d_zoom_large}. The circle center is then stepped in increments of 1 pixel along the straight line segments. When the difference between the $\phi$ and $\theta$ angles is the smallest, this indicates that the circle is at a location where the tracks are coming out straight from the center point, likely indicating a kink feature. The algorithm computes the  \textit{magnitude difference} $d\theta = |\theta - \phi|$ at each step and stores it per time tick. The stepping stops when the algorithm has scanned a 40x40 pixel region around the initial vertex seed. This procedure is repeated for each vertex seed in the plane. The evolution of $d\theta$ as a function of the corresponding time tick is represented in Fig. \ref{3d_zoom_large}. If two vertex seeds happen to occur on the same time tick, the lowest $d\theta$ value is stored.

\begin{figure}[!t]
\center
\begin{overpic}[trim = 0 0 0 6.27cm, clip, width=0.75\textwidth]{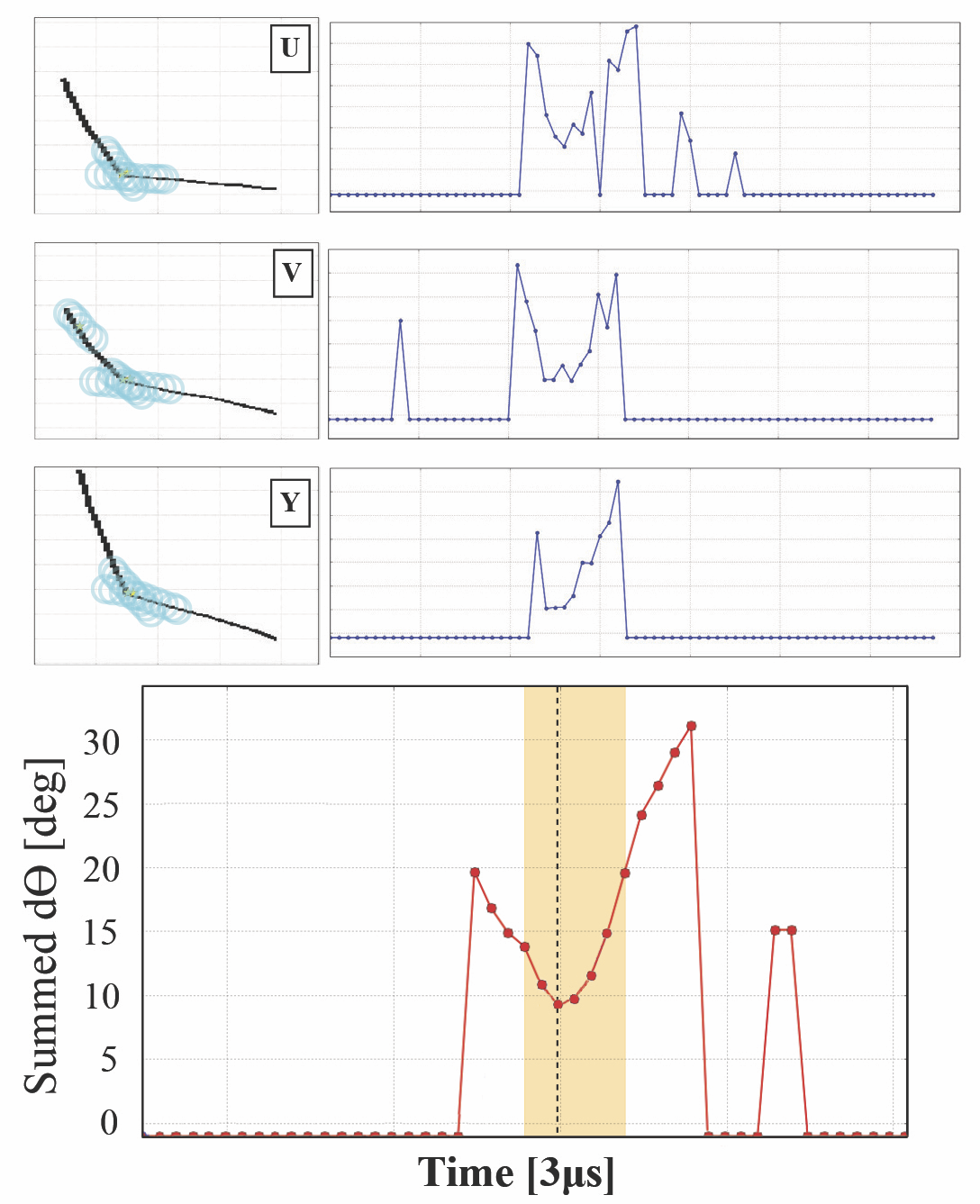}\put (14,54){\textbf{\scriptsize MicroBooNE Simulation}}\end{overpic}
\caption{
\label{3d_all_spectrum}
Determination of the best seed location across the three planes. The scanning procedure described in Fig. \ref{3d_zoom} is carried out for each vertex seed across three planes. The sum $d\Theta = \sum_{\textrm{plane}} d\theta$  is used to determine the best seed location in time. 
}
\end{figure}

Next, the $d\theta=|\theta-\phi|$ versus time-tick maps for the three views are summed to produce a single distribution of $d\theta$ values. The summed magnitude difference is the angular metric to be minimized to find the best vertex seed. The distribution is smoothed using a rolling mean that averages over 6 time ticks. Finally, the algorithm searches for local minima in the spectrum to find regions where a coincident vertex feature appears across multiple planes as shown in Fig. \ref{3d_all_spectrum}. In each plane, the circle at the local minima time is examined and matched across planes using wire coincidences. If coincident wires are found, then a 3D vertex is claimed. 
A set of coincident vertex seeds across three planes will exhibit a more pronounced local minimum in the summed $d\theta$ variable than a set of seeds that only coincide on two planes. Therefore, while it is possible to find vertex candidates that only match on two planes, a candidate that matches on three planes is favored.

To obtain the 3D vertex position,  the vertex time provides the $X$-coordinate. The $Y$ and $Z$ spatial information of the vertex is extracted from wire coincidences across any pair of $U$, $V$, and $Y$ planes.

\subsection{Refining the 3D-Vertex Position}

The algorithm then refines the 3D vertices found in both the track-only and the track-shower cases to best estimate the final 3D vertex. This procedure is called the ``3D vertex scan". The algorithm takes as input a 3D point in space, and scans in steps of \SI{0.5}{\centi \meter} a \hbox{($4\times 4 \times 4$) \si{\cubic \centi \meter}} volume around the 3D point. At each 3D step in the local volume, a 2D point is formed by projecting into the plane. At the projected step point, a circle is gradually increased in size from 6 pixels to 10 pixels in steps of 2 pixels. At each radial step the 2D angular metric (summed magnitude difference between $\theta$ and $\phi$) is computed. The result is summed across planes, and minimized over the 3D volume. The 3D points which minimize the angular metric are claimed as 3D vertex candidates. 

\subsection{Correlating Particles Across the Three Views}
Once the 3D vertex is identified, the images across the three views of each particle that emanates from the vertex must be correlated.  This is done in two steps.   First, in each view, unique particles are identified using 2D clustering.   Then, these 2D clusters are matched across views.

\begin{figure}[!t]
\center
\subfigure[]{\begin{overpic}[trim = 0cm 12cm 0cm 0cm, clip, width=0.7\textwidth]{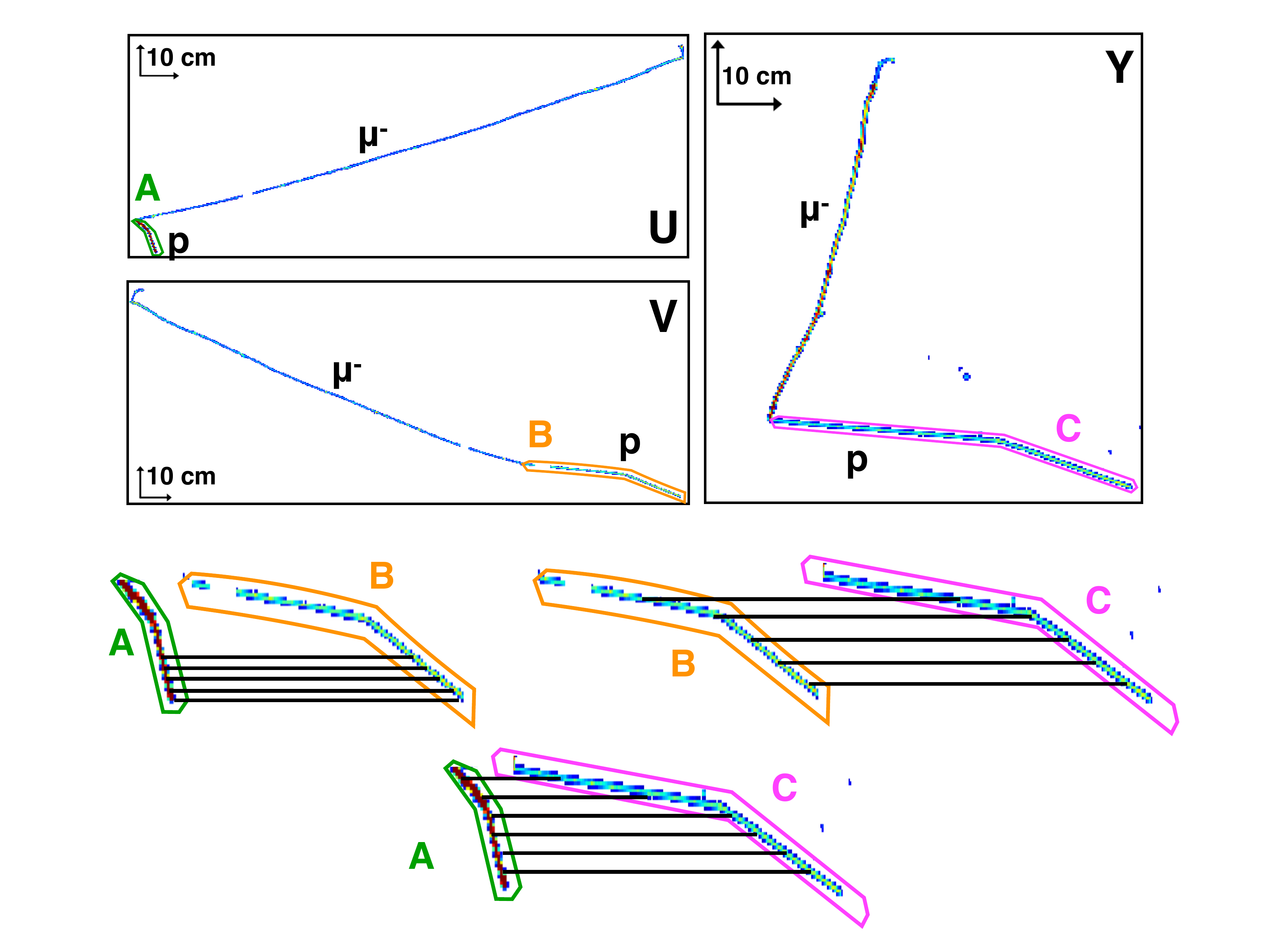}\put (10,40){\textbf{\scriptsize MicroBooNE Simulation}}\end{overpic}}
\subfigure[]{\begin{overpic}[trim = 0cm 0cm 0cm 15cm, clip, width=0.7\textwidth]{vtx_algo/numu_track_match.pdf}\put (10,40){\textbf{\scriptsize MicroBooNE Simulation}}\end{overpic}}
\caption{
\label{matching_diagram}
A cartoon matching scenario for a triplet of clusters across the U,V, and Y planes. (a) A green (A), orange (B), and pink (C) cluster reconstructed from a particle vertex for each of the three plane views. (b) Triplet pairings for possible cluster combinations. For each combination, pixels in each cluster are compared for 3D consistency via time tick and wire overlap (black lines). The ratio of the number of 3D consistent pixels to the number of pixels in the smallest cluster is computed for each combination. 
}
\end{figure}

In this step, 2D clusters at each vertex candidate point are matched across planes to identify a unique 3D particle using the ADC image. The clusters from a single vertex are matched by first computing all possible combinations of clusters across 2 and 3 planes from a given vertex. For each combination of 2D clusters, a score is computed. This score is a floating point number between 0 and 3 and is calculated in the following way. There are two cases to consider: a two-plane match (pairing) and a three plane-match (triplet). For a two-cluster pair, the smallest and largest clusters are determined based on their total number of pixel points. For each pixel in the smallest cluster, the corresponding coincident pixel in terms of time tick (X), and wire overlap (Y,Z) position is searched for in the larger cluster. Wire overlap refers to wires in two different planes that, for a given time tick value, are expected to show charge deposition according to a (Y,Z) position in the detector. The total coincident overlap between the larger and smaller cluster is computed as the ratio of the number of overlapping pixels in the smaller cluster, to the total number of pixels in the smaller cluster. The pair score will be a number between 0 (no overlap) and 1 (the smallest cluster is covered completely). A threshold of 0.5 is applied to ensure the potential match is unique. For the three plane match, the same procedure is calculated for each pair of clusters in the triplet. A threshold  overlap value of 0.5 is applied between combinations to ensure a unique match. The triplet score is determined from a sum of overlaps across the triplet. If at least 2 clusters in the triplet do not meet the threshold requirement, the match is discarded. The triplet score will be a number between 0 (no overlap) and 3 (all three clusters overlap completely). In this way, a three plane match is favored over a two plane match since the potential for a score larger than 1 is possible. For each vertex the algorithm returns unique combinations of matched clusters by comparing the scores. For each match, a 3D particle candidate is identified. An illustration for the triplet case is shown in Fig. \ref{matching_diagram}.

\subsection{3D Vertex studies}
\label{subsec:mcvtxeffstudy}

In this section, we discuss the quality of the vertex reconstruction.   We consider the resolution of the 3-D vertex position reconstruction and the efficiency of vertex finding for the MicroBooNE detector.

For this study, we use a simulated $\nu_\mu$ sample filtered to satisfy our target final state choice ``$1\mu1p$''. The simulation is described in detail in Section \ref{intoPaper}. The true vertex is required to be within a fiducial volume defined by  \SI{10}{\centi \meter} from any edge of the active volume. Also, we require that the event contains only one muon and one proton, having a kinetic energy greater than \SI{35}{\mega \electronvolt} and \SI{60}{\mega \electronvolt}, respectively at the generator level, and that the muon is contained in the active volume. $1\mu 1p$ events in MicroBooNE that satisfy the containment criterion typically have energies that range from \SI{200}{\mega \electronvolt} to \SI{2}{\giga \electronvolt}.

\subsubsection{Vertex Resolution}

\begin{figure}[!t]
\center
\begin{overpic}[width=0.75\textwidth]{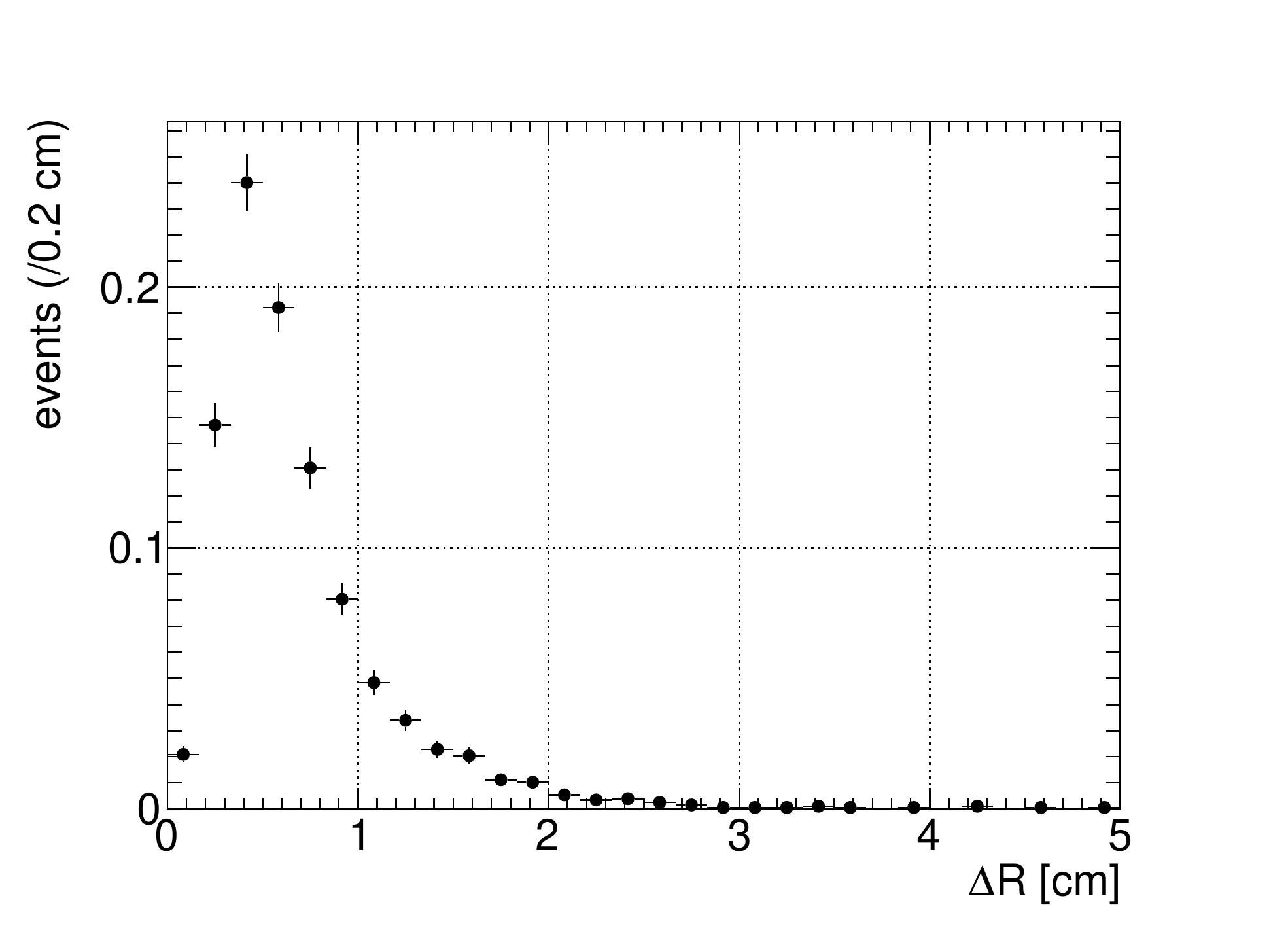}\put (13,64){\textbf{\scriptsize MicroBooNE Simulation}}\end{overpic}
\caption{
\label{Track3Dres} Vertex resolution for reconstructed $1\mu1p$ events. The 3D distance, $\Delta R$, between the predicted true neutrino vertex and the reconstructed vertex is shown.  Of the reconstructed vertices, $68 \%$ are found within $0.73$\si{\centi \meter} of the true vertex position. The true neutrino vertex position has been corrected for displacements due to space-charge effects.}
\end{figure}

The quality of the track image vertex-finding can be assessed using MC by considering $\Delta R$, the distance between the simulated true neutrino vertex and the reconstructed vertex. The $\Delta R$ distribution is shown in Fig. \ref{Track3Dres}, unity normalized. Electric field inhomogeneities throughout the detector volume can cause distortion of tracks and displacement of the apparent vertex location. This effect is known as the space-charge effect \cite{Mooney:2015kke}. To compensate for this, and in order to estimate the resolution of the vertex reconstruction, we correct for the space charge distortion introduced in the simulation before estimating $\Delta R$. Of the reconstructed events, $68 \%$ have their vertex found within  $0.73$ \si{\centi \meter} of the true vertex position.

\subsubsection{Vertex-finding Efficiency}
\label{vertexEff}

\begin{figure}[!t]
\centering
\begin{overpic}[width=0.75\textwidth]{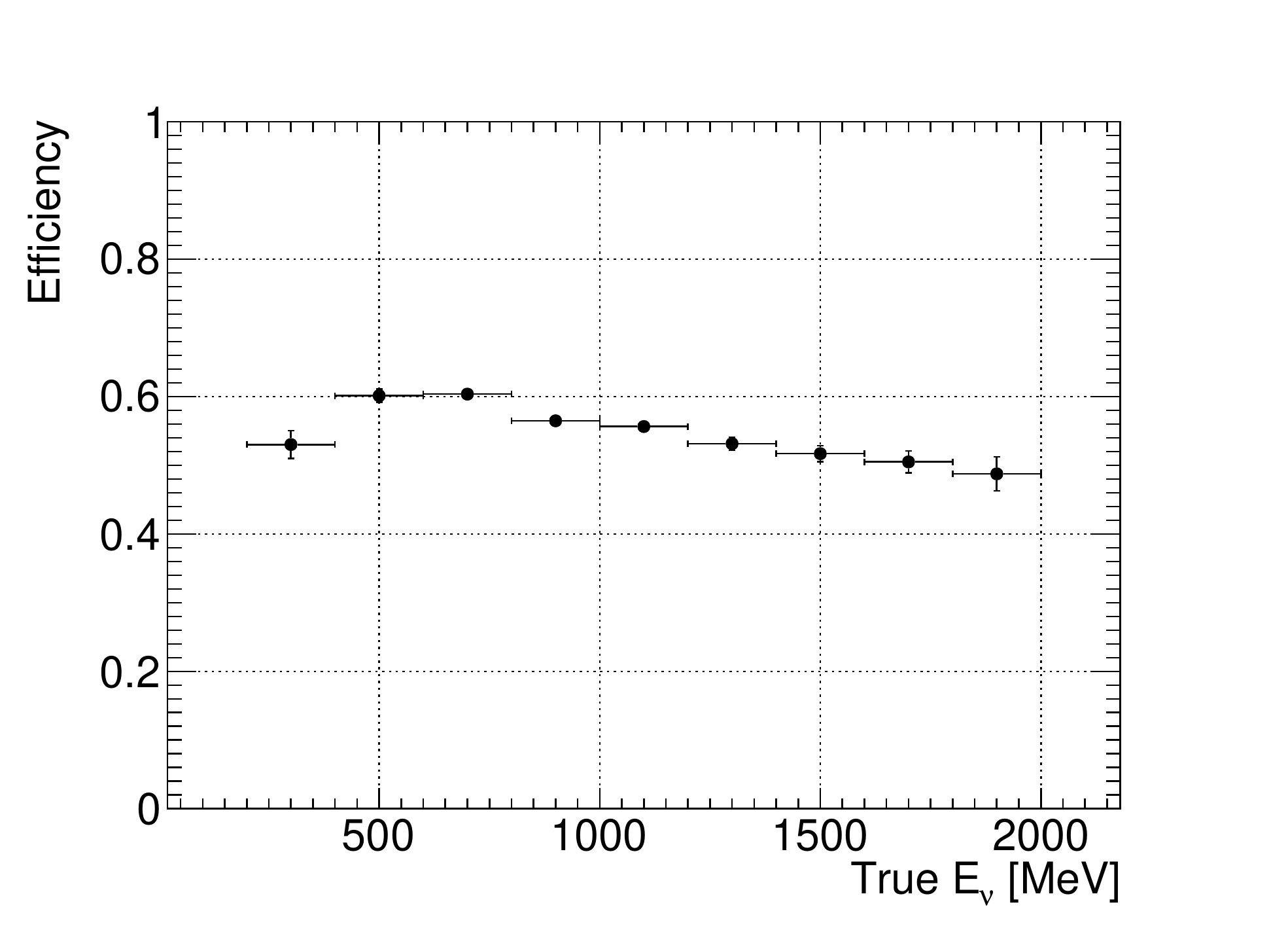}\put (13,64){\textbf{\scriptsize MicroBooNE Simulation}}\end{overpic}
\caption{\label{fig:eff_all_energyInit} Overall vertexing efficiency as a function of true neutrino energy for $1\mu1p$ events for which the true vertex is included in the found cROI.}
\end{figure}

Figure \ref{fig:eff_all_energyInit} shows the efficiency of finding a vertex for a $1\mu 1p$ $\nu_\mu$ event for which the true vertex is within a found cROI as a function of the true neutrino energy. Well-reconstructed  events are defined as events with at least one reconstructed vertex within 3 cm of the true vertex position after correcting for space charge effects.  An average efficiency for the vertex finding algorithm of $56\%$ is found.  Lower efficiencies are found for for low and high energy events. Low energy events tend to be harder to reconstruct due to the smaller extent of at least one of the particle tracks exiting the vertex. Two-track clusters of at least 6 pixels need to be found and correlated through at least two planes for a vertex to be reconstructed; a low energy event is less likely to pass this requirement. Conversely, at higher energies, the opening angle between the two tracks tends to be smaller; it is then more likely that due to projection effects, the track clusters overlap in at least one plane, which in conjunction to unresponsive wires, can lead to a lower vertex finding efficiency.

\begin{figure}[!t]
\centering
\vspace{1 cm}
\begin{overpic}[trim = 0.8cm 0 0.8cm 14cm, clip, width=0.75\textwidth]{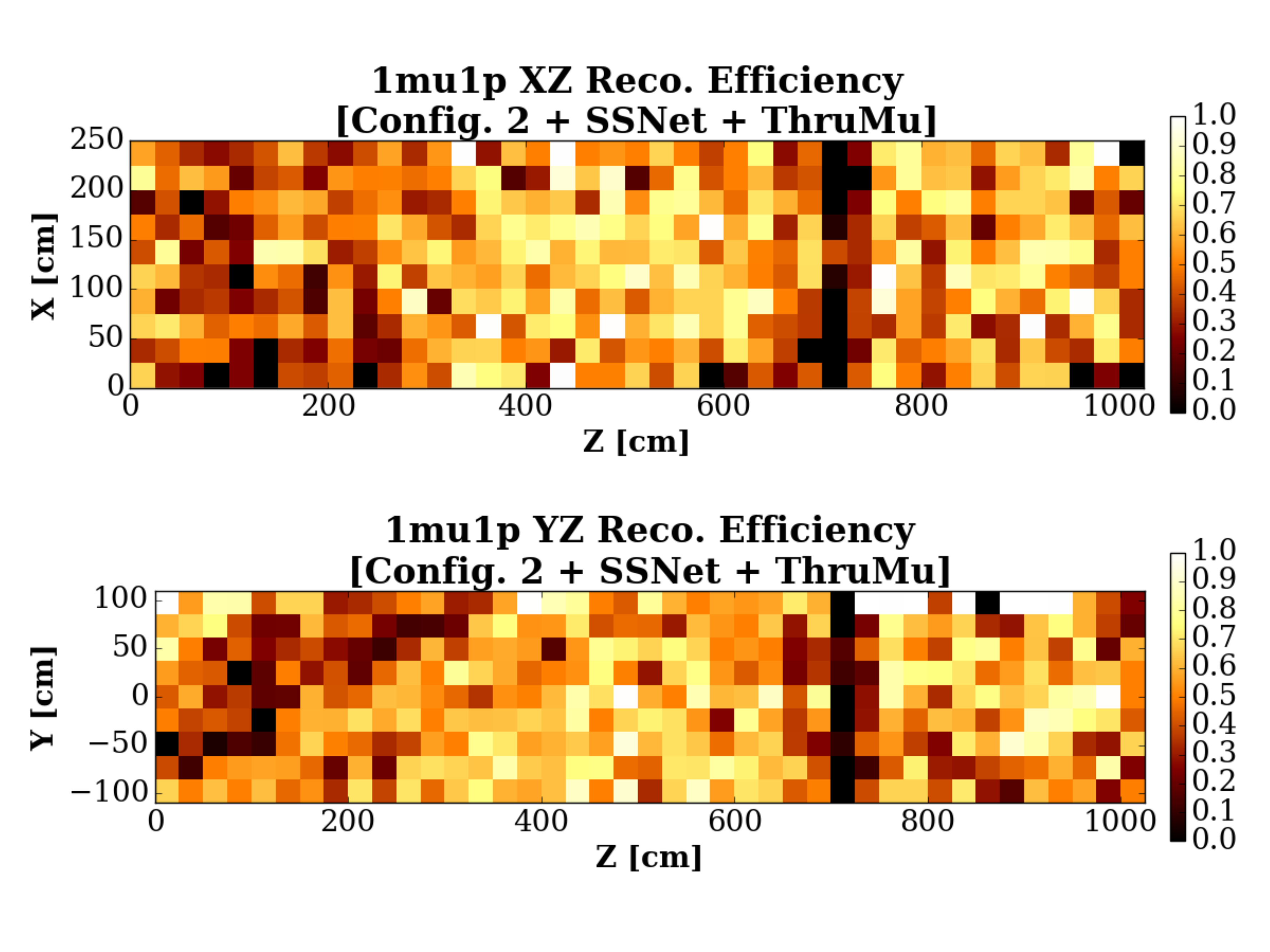}\put (11,37){\textbf{\scriptsize MicroBooNE Simulation}}\end{overpic}
\caption{\label{fig:numu_all_croi_eff} Efficiency for finding a well-reconstructed $1\mu1p$ vertex in the (Y,Z) plane. Each cell is the ratio of the number of events with a well-reconstructed vertex to the number of neutrino events in that cell.}
\end{figure}

Figure \ref{fig:numu_all_croi_eff} shows the spatial dependence of the vertex finding efficiency when applying all upstream stages of the reconstruction. Two major unresponsive regions can be identified that have a noticeable impact on the vertexing algorithm. There is a band shown on the Z plane around \SI{700}{\centi \meter} on the Z axis where no reconstructed vertex is found. In addition, there is a low efficiency band starting at a Y value of approximately \SI{-100}{\centi \meter} sloped upward to a Z value of approximately \SI{200}{\centi \meter}. Both these regions are consistent with two known unresponsive regions, one on the Y plane, and one on the U plane, within the MicroBooNE detector.

\section{3D Track Reconstruction Algorithm}
\label{sec:astaralgo}

\subsection{3D track finding}

The reconstruction of 3D tracks is required to obtain the track kinetic energy.  We convert the 3D length of the track to kinetic energy using the known stopping power of each type of particle in liquid argon \cite{NIST} \cite{PSTAR}.   The full track length cannot be accurately inferred from the 2D images; thus 3D reconstruction is necessary.   Particle identification is also necessary in order to apply the appropriate stopping-power conversion.

Track reconstruction is particularly sensitive to the quality of the image on a large scale, and to data-Monte Carlo differences.   An interruption of the charge deposition along the track, due to unresponsive or noisy wires, waveform deconvolution artifacts, etc., may lead to an incorrect misreconstructed length and ultimately to the reconstruction of an inaccurate energy.  

Because of the sensitivity of the OpenCV-based pixel clustering to unresponsive wires and SSNet labeling continuity, only the position of the vertex is used in the track reconstruction, not the particle-by-particle clustering. Moreover, the location of the vertex within the cROI may cause the tracks and showers to exit that cROI. If this happens, although the vertex can be identified, the particle clustering may not reach the end of the tracks and showers, leading to misreconstructing them. It is important to have the capability to pursue a reconstruction beyond the cROI.
Searching for the 3D path followed by the track all the way to the end of the track is thus performed by a separate tracking algorithm.\\

The track finding algorithm takes as input the full ADC images for each of the three planes and the vertex 3D point. No other information is used. 
As unresponsive wires can change in time, it is then important to identify them for each event. The detection of unresponsive wires is achieved by tagging the wires over which no charge is deposited for the full read out duration of the event.\\

The reconstruction of a track begins by finding a set of 3D points, i.e. points in the ($x$,$y$,$z$) coordinate frame of the cryostat, that belong to a given track by performing an iterative stochastic search in the neighborhood of previously found 3D points.
A regularization is then performed to find a minimal set of ordered 3D points that describes the track at the required spatial resolution.
Finally, observables such as length, local and average charge deposition, and angles can be estimated.

For finding the first set of 3D points, the algorithm proceeds in successive reconstruction steps starting at the vertex:
\begin{itemize}
\item The seed of each reconstruction step  (iteration) is placed at the last found point.
\item A set of random 3D points is picked inside a sphere, centered at the seed, of radius $r_{\mathrm{search}} = 2+4 \times e^{(-L/5)}$ \si{\centi \meter}, where $L$ represents the distance between the vertex and the last points already found in that track. The purpose here is to allow a wider search around the vertex point, but once a track is found, characterized by a significant $L$ of \SI{5}{\centi\meter}, the restricted radius helps to prevent the track reconstruction from jumping to a nearby track. The search region rapidly decreases to \SI{2}{\centi\meter}, while tracks are typically $\approx$ \SI{1}{\centi\meter} wide. This allows coverage of the whole thickness of the tracks and good coverage when crossing other tracks.
\item If the current track is longer than \SI{5}{\centi \meter}, 3D points are also selected inside a forward-going cone within a $30^{\circ}$ opening angle. The cone follows the averaged direction of the track in the last \SI{10}{\centi \meter}, or the full length of the track if it is too short. The forward search region extends to $2 \times r_{\mathrm{search}}$. While the spherical search region allows for fine tracking, and can resolve sudden direction changes, the forward cone search region allows for a faster progression on straight track portions and increases the robustness to unresponsive wires.
\item Only the points that project back onto pixels with non-zero charge deposition on all planes are kept, with at most one plane on which the point projects on an unresponsive wire.
\item New points can only be added if the sum of the ADC values of the deposited charge on the pixels on which they project is greater than that of the already placed points of that iteration.
\end{itemize}

At this point, we have a set of neighbors to the seed. Some of these points are not relevant, because they are too close to an already found point or track.  They are rejected based on the following criterion:

\begin{itemize}
\item New points cannot be placed closer than \SI{0.3}{\centi \meter} from an already placed point.
\end{itemize}

All the remaining points at this stage are added to a proto-track, a cloud of unrelated 3D points that correspond to non-zero pixels. The point within the set newly found points that is the furthest away from the current seed is now used as the new seed and that phase is iterated as long as new points can be found. This phase ensures that the explored region is pushed as far as possible along the track.\\

\begin{figure}[!h]
\center
\includegraphics[width = 0.45\textwidth]{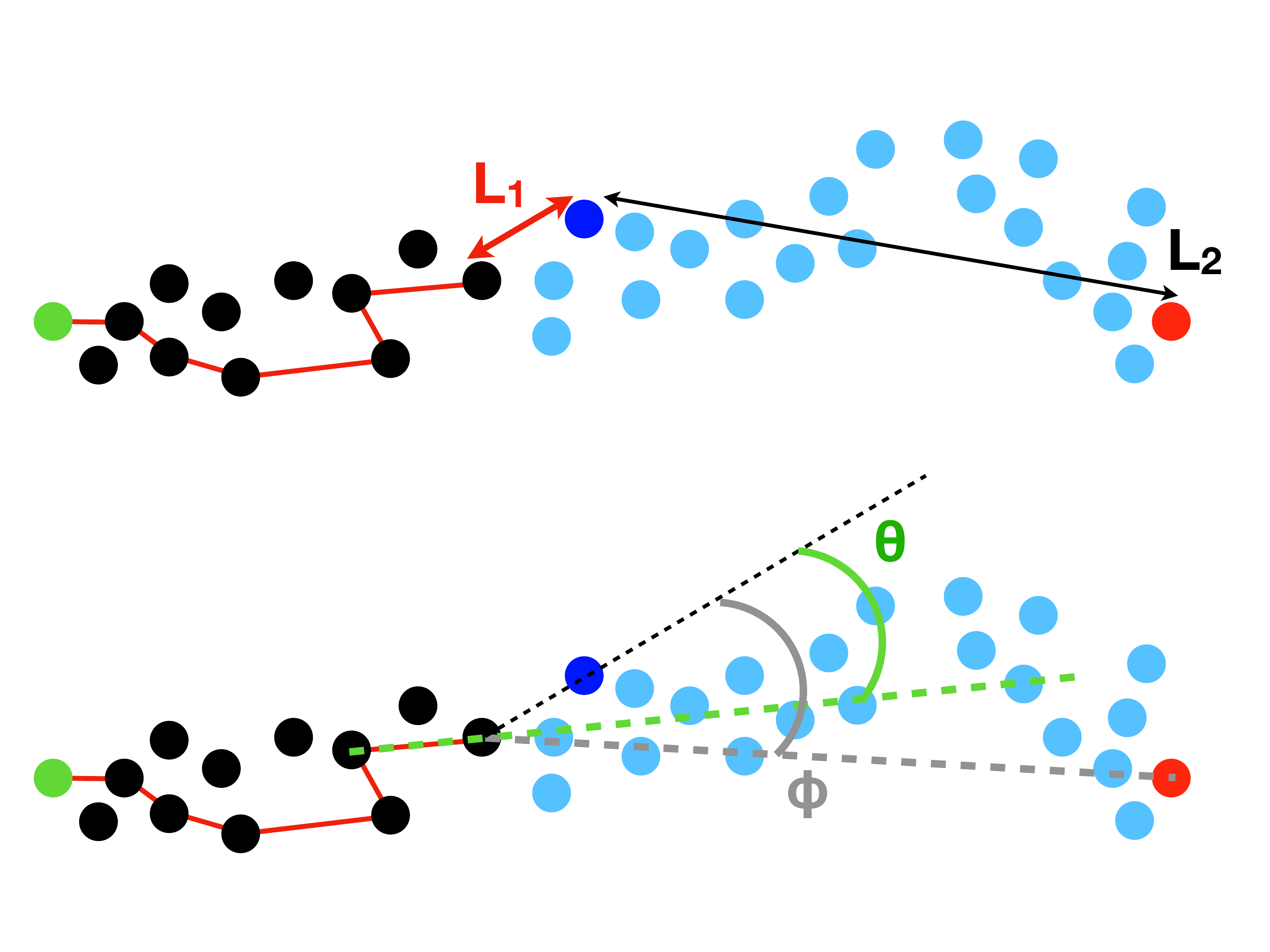}
\caption{\label{SortNorderPts} Cartoon of the 3D points found in the first stage of track reconstruction. These points correspond to 3D locations consistent with charge deposition on all three planes, but have no logical link between them. Once the set of uncorrelated 3D points is found, a sorting algorithm finds a logical path. From a sorted point, the other candidates (dark blue point for current step) are evaluated based on the distance to already sorted points (black points), the remaining distance to the end of the track (red point) and the two angles, with respect to the last sorted points ($\theta$) and to the end of the track ($\phi$). The green point represents the vertex and the light blue points are the points of the found set that have not been sorted through yet.}
\end{figure}

The points in the proto-track are not ordered and do not follow a linear path. They zig-zag back and forth within the thickness of the track. The next step is to order the points by linking each one to its most likely neighbor. This next neighbor is found by minimizing a global score over all unordered points:
\begin{align}
\textnormal{global score} =~ &5 \cdot L_{1} + 0.1 \cdot L_{2}\\
 			               +~ &2\cdot (2- \cos \theta) \notag\\
 			               +~ &10\cdot (2- \cos\phi).  \notag 
\end{align}
The distances $L_{1}$ and $L_{2}$ (in centimeters), as well as the angles $\theta$ and $\phi$ are summarized in the cartoons in Fig. \ref{SortNorderPts}.
The dots represent the set of 3D points in the proto-track. The black dots are the points that have been sorted through, and the light blue dots are the remaining un-sorted points. The green and red dots correspond to the vertex, and the end of the track, respectively. The end point is selected as the 3D point which is the furthest away from the vertex. The red broken line corresponds to the path found within the sorted 3D points. For each candidate within the points that are still un-sorted (the dark blue point in Fig. \ref{SortNorderPts} shows one) the two lengths and angles are computed : $L_{1}$ is the distance to the last selected 3D point, $L_{2}$ is the distance to the end of the proto-track, $\theta$ is the angle from the last two sorted points to the candidate, and $\phi$ is the angle between the candidate, the last sorted point, and the end of the proto-track.  Once the points in the proto-track have been ordered, there is a logical path from one point to the next, and some points are rejected as they are never the best candidates, they are then discarded from the list of available points. The numerical parameters in the global score have been tuned to optimize the smoothness of the paths, while allowing it to bend and follow the particle path with fidelity. However, at this point, the track still zig-zags and is formed by too many 3D points to be a good representation of the particle path, so a second stage is required.\\
 
The second stage smooths the path. We loop through the set of 3D points, rejecting superfluous points based on several criteria:

\begin{itemize}
\item A new set of points is created by performing a rolling average of two consecutive points.
\item The new set is ordered by moving from one point to its closest neighbor.
\item The selected new point must be closer to the end point than the previous one.
\item The distance from the previous point cannot be more than \SI{5}{\centi \meter} (as this would indicate a possible jump to another nearby track).
\item The points that deviate by less than \SI{0.5}{\centi \meter} from the line between points n-1 and n+1 are removed.
\end{itemize}


\subsection{Full Event Reconstruction}

These operations are then iterated for all tracks that are found.   
To prevent the algorithm from finding the same track multiple times, or from producing equivalent tracks that explore the same actual charge deposition, the pixels corresponding to a found track are masked in the ADC image. Two regimes are used to mask the pixels:
\begin{itemize}
\item The 3D points are within \SI{2}{\centi \meter} of the vertex: pixels within a 3-pixel cylinder around the projected track on each plane are erased.
\item The 3D points are beyond \SI{2}{\centi \meter} of the vertex: pixels within a 6-pixel cylinder around the projected track on each plane are erased.
\end{itemize}
Pixels are erased on a smaller cyliinder close to the vertex in order to allow the algorithm to be efficient at finding tracks that overlap, i.e. that would have a small projected angle with one another, in one of the three planes.\\

Once new tracks are no longer found, we iterate the process to the end points of the tracks already found. The end points were selected as the point the furthest away from the vertex, but in some cases, if scattering causes the track to curl up, the actual end of the track is not the furthest point. Starting at the end of a found track and looking for a missing portion of the track helps correct for these cases.  The two portions of the same tracks are then merged into a single new track.

\subsection{Self-Diagnostic}

\begin{figure}[!t]
\center
\subfigure[\label{diagnoseU}Reconstruction ends in unresponsive wires]{ \begin{overpic}[trim = 0 23cm  0  0, clip, width = 0.43\textwidth]{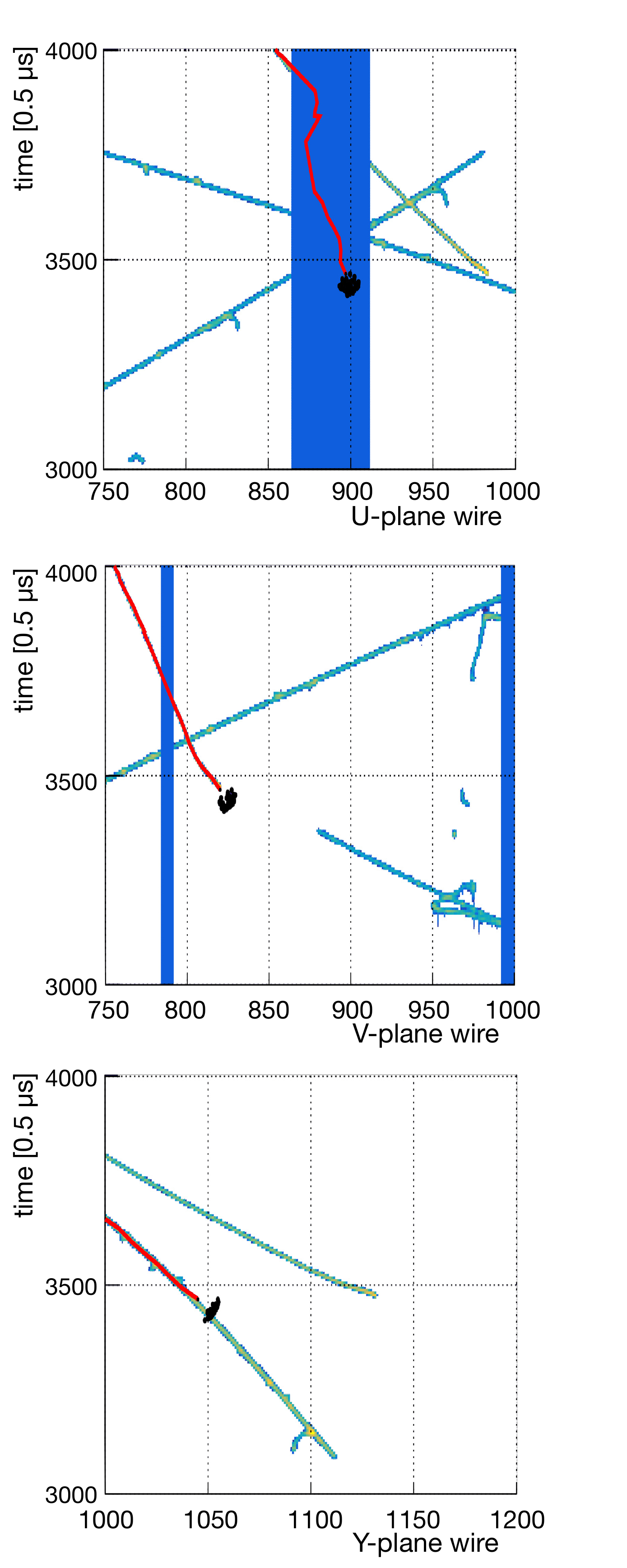}\put(16,82){\textbf{\scriptsize MicroBooNE Simulation}}\end{overpic}}
\subfigure[\label{diagnoseV}Reconstruction ends in empty region]{\includegraphics[trim = 0 11.4cm  0  12.5cm, clip, width = 0.43\textwidth]{figs/p2ss_3drecoalgo/Tracks_00001_00101_90317_0011_3.pdf}}
\subfigure[\label{diagnoseY}Reconstruction ends while track continues]{\includegraphics[trim = 0  0  0  24cm, clip, width = 0.43\textwidth]{figs/p2ss_3drecoalgo/Tracks_00001_00101_90317_0011_3.pdf}}

\caption{\label{endPointCharacterization}Example of a simulated event that exhibits the three possible ends to a reconstructed track : (a) the track ends while on an unresponsive region, (b) the track ends at the termination of the charge deposition on that plane, and (c) the track ends while the charge deposition continues on that plane. The red line corresponds to the path of the reconstructed track, projected on each plane, and the black dots represent the search regions to diagnose the end of the track. The track end point labels of these three cases are Unresponsive (a), Empty (b) and Track  (c).}
\end{figure}

Once all the tracks associated with a vertex have been found, it is important to recognize and possibly reject cases where the reconstruction failed. The purpose of the following step is to identify why no more points can be associated to the found tracks, either because the reconstruction reached the end of the actual particle path, or because an error, or an unresponsive region, occurred, and no more 3D point can be projected into non-zero pixels in all three images.
This cross-check relies on a set of random points thrown on a spherical shell of radius \SI{3}{\centi \meter} at the end point of each track. Only the forward-going points within a solid angle of $65^\circ$ are kept, where forward is defined as the direction along of the track.
For each track, the fraction of points that project onto pixels corresponding to unresponsive wires, empty pixels and pixels with charge deposited are evaluated, and a label is attributed to the end point on each plane.
Figure \ref{endPointCharacterization} shows the three possible cases we distinguish. The red line corresponds to the projected reconstructed path on a given wire plane of the reconstructed 3D points, the colored pixels are the charge depositions, and the uniform blue region corresponds to unresponsive wires. The black dots correspond to the points used to estimate the goodness of the reconstruction. 


In the case where the charge deposition seems to continue (end point label ``Track"), it is clear that the reconstruction failed to find the actual end of the track, but in the cases where the track ends at the termination of the charge depositions (end point label ``Empty") or on an unresponsive region (end-point label ``Unresponsive"), one needs to have a more global approach and look at the other planes.\\

A track that reaches the end in an unresponsive region in two planes but in the third plane shows that the end of the charge depositions is reached, can be considered as well reconstructed.
However, the case where the charge deposition seems to continue in the third plane indicates that the reconstruction did not reach the end of the track.

Another case that can be addressed in this way is when the track locally becomes coplanar to a wire in the induction planes. Since the signal in these wires is bipolar, the track can appear faint, or interrupted. In that case, the reconstruction will sometimes stop at the interruption point. This end point is going to be classified as correct since it appears that, indeed, the track has stopped. In the other two planes, however, the end points will be classified as bad because the charge deposition seems to continue.

Based on the label of the end points in the various planes, a track label is attributed to the tracks as shown in Table \ref{VertexCharacterization}.
\begin{table}[!h]
\caption{\label{VertexCharacterization}Track label attribution for the three possible outcomes of the diagnostic tool, depending on the number of planes whith a given end point label.}
\center

\begin{tabular}{|c|c|c|c|}
\hline
\multicolumn{3}{|c|}{Number of planes} &\multirow{3}{*}{Diagnostic Output}\\
\multicolumn{3}{|c|}{with a given end label} &\\
\cline{1-3}
Unresponsive & Empty & Track & \\
\hline
3 & 0 & 0 & Unresponsive Wire\\
2 & 0 & 1 & Unresponsive Wire\\

1 & 1 & 1 & ~Failed Reconstruction~\\
0 & 2 & 1 &  ~Failed Reconstruction~\\
0 & 1 & 2 &  ~Failed Reconstruction~\\
0 & 0 & 3 &  ~Failed Reconstruction~\\
1 & 0 & 2 &  ~Failed Reconstruction~\\

1 & 2 & 0 & Good Reconstruction\\
2 & 1 & 0 & Good Reconstruction\\
0 & 3 & 0 & Good Reconstruction\\
\hline
\end{tabular}

\end{table}

\begin{figure}[!t]

\subfigure[]{\begin{overpic}[trim = 0 23cm  2cm  0, clip, width = 0.32\textwidth]{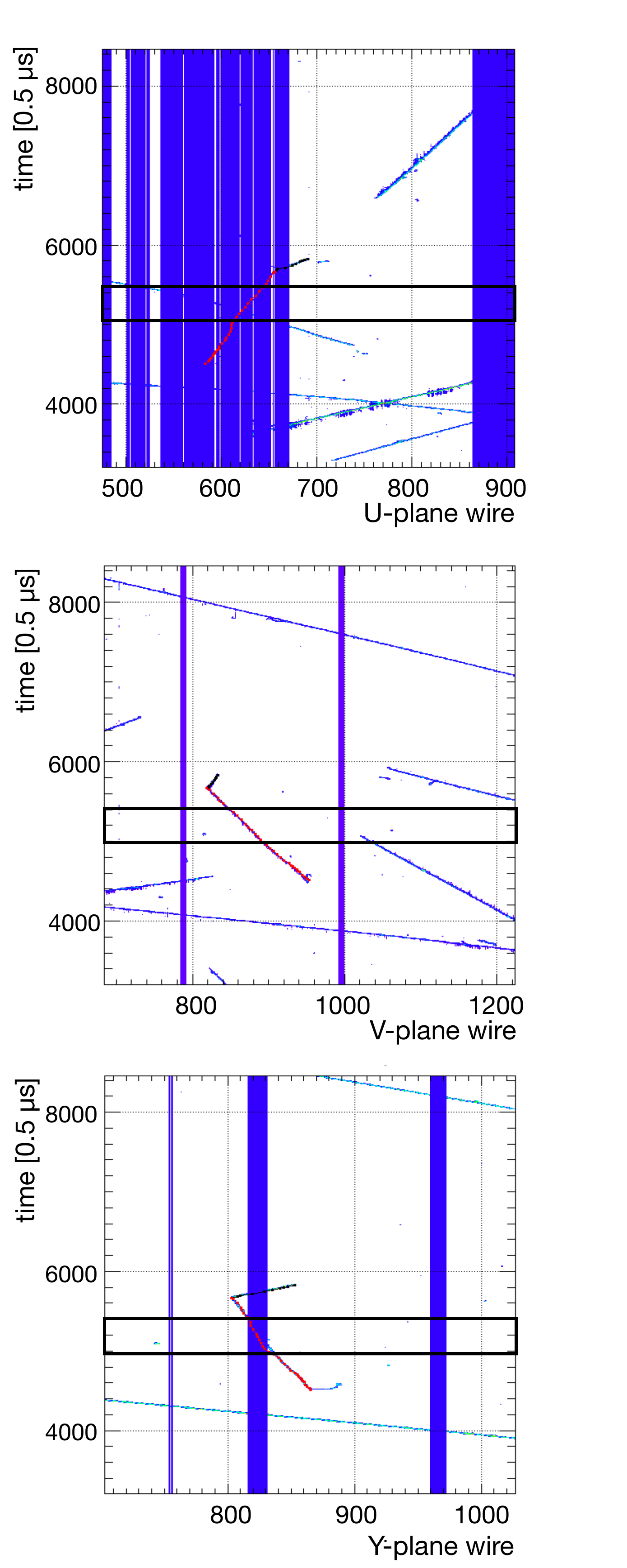}\put(19,92){\textbf{\scriptsize MicroBooNE Simulation}}\end{overpic}}
\subfigure[]{\includegraphics[trim = 0 11.4cm  2cm  12.5cm, clip, width = 0.32\textwidth]{figs/p2ss_3drecoalgo/cVertex_07001_00115_05793_0000_0000__OKtracks_2.pdf}}
\subfigure[]{\includegraphics[trim = 0  0  2cm  24cm, clip, width = 0.32\textwidth]{figs/p2ss_3drecoalgo/cVertex_07001_00115_05793_0000_0000__OKtracks_2.pdf}}

\caption{\label{deadWireRecovery}Example of track reconstruction, where the track crosses unresponsive regions. In this example, the muon track (red) crossed unresponsive regions (vertical blue lines) in two planes, (a) and (c) at the same time (black rectangles) and would have been labeled as failed. The recovery method allowed a new stage of track finding to successfully reconstruct the track.}
\end{figure}

Cases that fail due to unresponsive wires can be recovered in some cases, as illustrated in Fig. \ref{deadWireRecovery}: Non-zero pixels are added in unresponsive regions along a line extrapolated from the direction of the track at the failure point. The track reconstruction and self diagnostic are then performed again. In cases where the tracks are reasonably straight within the unresponsive region, the track finding is able to connect the track to the opposite end of the unresponsive regions through the added pixels. After applying this correction, $0.3\%$ of events have track reconstructions terminated with the label "Unresponsive Wire", $38\%$ are terminated with the label "Failed Reconstruction", and $61\%$ with the label "Good Reconstruction".\\

In the following sections of the paper, a ``well-reconstructed vertex" is a vertex with exactly two tracks of more than \SI{5}{\centi \meter} that both end at the termination of the charge deposition.\\

From the reconstructed 3D-path of each particle exiting a vertex, several key observables can be estimated.  As we describe these observables in the subsections below, we will characterize the results using a $1\mu1p$ MC sample. In this sample, all $\nu_\mu$ events are generated with exactly one proton above \SI{60}{\mega \electronvolt} of kinetic energy and exactly one muon above \SI{35}{\mega \electronvolt} of kinetic energy, and the neutrino interaction is a quasi-elastic charged current interaction. A fiducial volume defined by \SI{10}{\centi \meter} from any edge of the active volume is  applied on the vertex position, and a containment criterion is applied to the two particles. Simulated neutrino interactions are overlaid onto a cosmic ray background image from unbiased beam-off data. Furthermore, in order to evaluate the performance of the track-finding algorithm specifically, we require that the vertex found is within \SI{2}{\centi \meter} of the true vertex position, and we compute the performance of the track finding algorithm (resolutions, efficiencies, etc.) on the events that follow this criterion. Furthermore, for the resolution studies in sections \ref{ChargeReco}, \ref{AngleReco}, and \ref{Ereco}, the events used were all labelled as well reconstructed by the self diagnostic tool. However, the efficiencies estimations in section \ref{EffReco} do include bad reconstruction events in the denominator of the efficiency, but not in the numerator.

\subsection{Local charge deposition \label{ChargeReco}}
For each 3D point, the local charge deposition ($dQ/dx$) for a given plane is computed by integrating the values of non-zero pixels in a 2 pixel radius around the projection of the 3D point on that plane. 

Once the local charge deposition around each 3D point has been acquired for each of the three planes, other variables can be derived at the analyzer's preference, including:
\begin{itemize}
\item \textbf{Collection-plane-only average ionization}: the averaged ADC value per 3D point 
\begin{align}
\left \langle \dfrac{dQ}{dx} \right \rangle &= \dfrac{1}{n}\sum_{i<n} \left(\dfrac{dQ}{dx}\right)_{i}
\end{align}
where n is the number of reconstructed 3D points and $\left(\dfrac{dQ}{dx}\right)_{i}$ is the local ionization of a given point $i$ using only the collection plane information.

\item \textbf{Three-plane average ionization}: the same variable as previously defined but using, for each point, the averaged pixel value of all three planes with non-zero charge. In this case, a scale factor $3/N$ is applied where $N$ is the number of planes on which a non-zero value was found. This scale factor allows correction for a possible plane in which the 3D point projects onto an unresponsive region.

\end{itemize}

\begin{figure}[!t]
\center
\begin{overpic}[width = 0.75\textwidth]{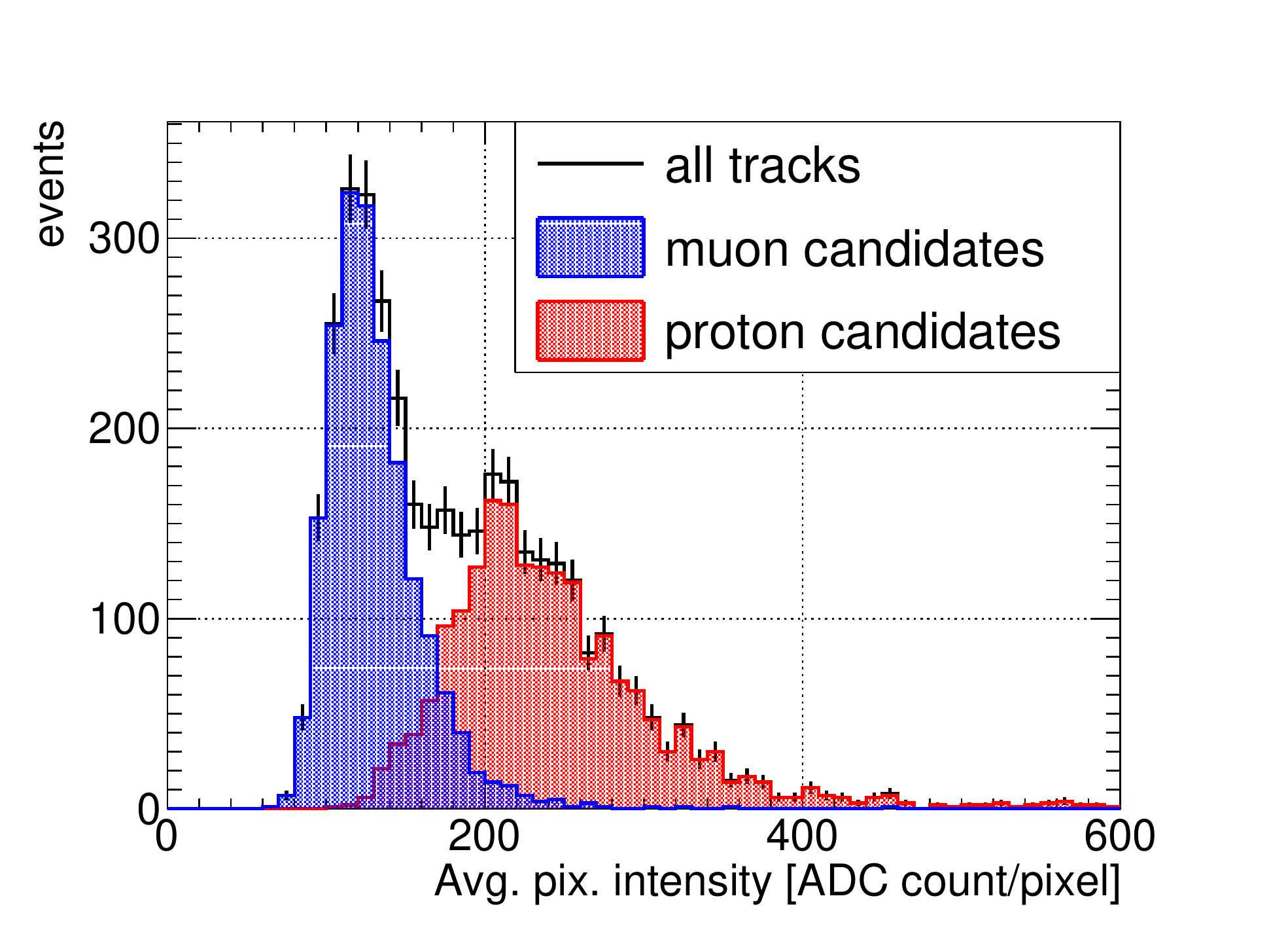}\put(13,65){\textbf{\scriptsize MicroBooNE Simulation}}\end{overpic} 
\caption{\label{AverageIonization} Pixel intensity averaged along the reconstructed track on the three planes. The red and blue distributions represent the tracks with the highest and lowest average ionization in a given vertex respectively. In the case of a $1\mu 1p$ $\nu_\mu$ interaction, the track with the higher pixel intensity is likely to be the proton.}
\end{figure}

The average ionization per pixel reconstructed for simulated $1\mu 1p$ $\nu_\mu$ events in MicroBooNE is shown in Fig. \ref{AverageIonization}. At this stage, no particle identification has been performed, the two populations have been separated by identifying the muon as the track with the lowest average ionization within the pair of reconstructed tracks (blue distribution) and identifying the proton as the track with the highest average ionization (red distribution). It is important to note that these particle identifications are relative within a pair of reconstructed particles, assuming one is a proton and the other a muon.  This is sufficient at this stage as a more definitive particle identification will be performed later on in the analysis chain.

\subsection{Angle Estimation \label{AngleReco}}

\begin{figure}[!t]
\center
\includegraphics[width = 0.7\textwidth]{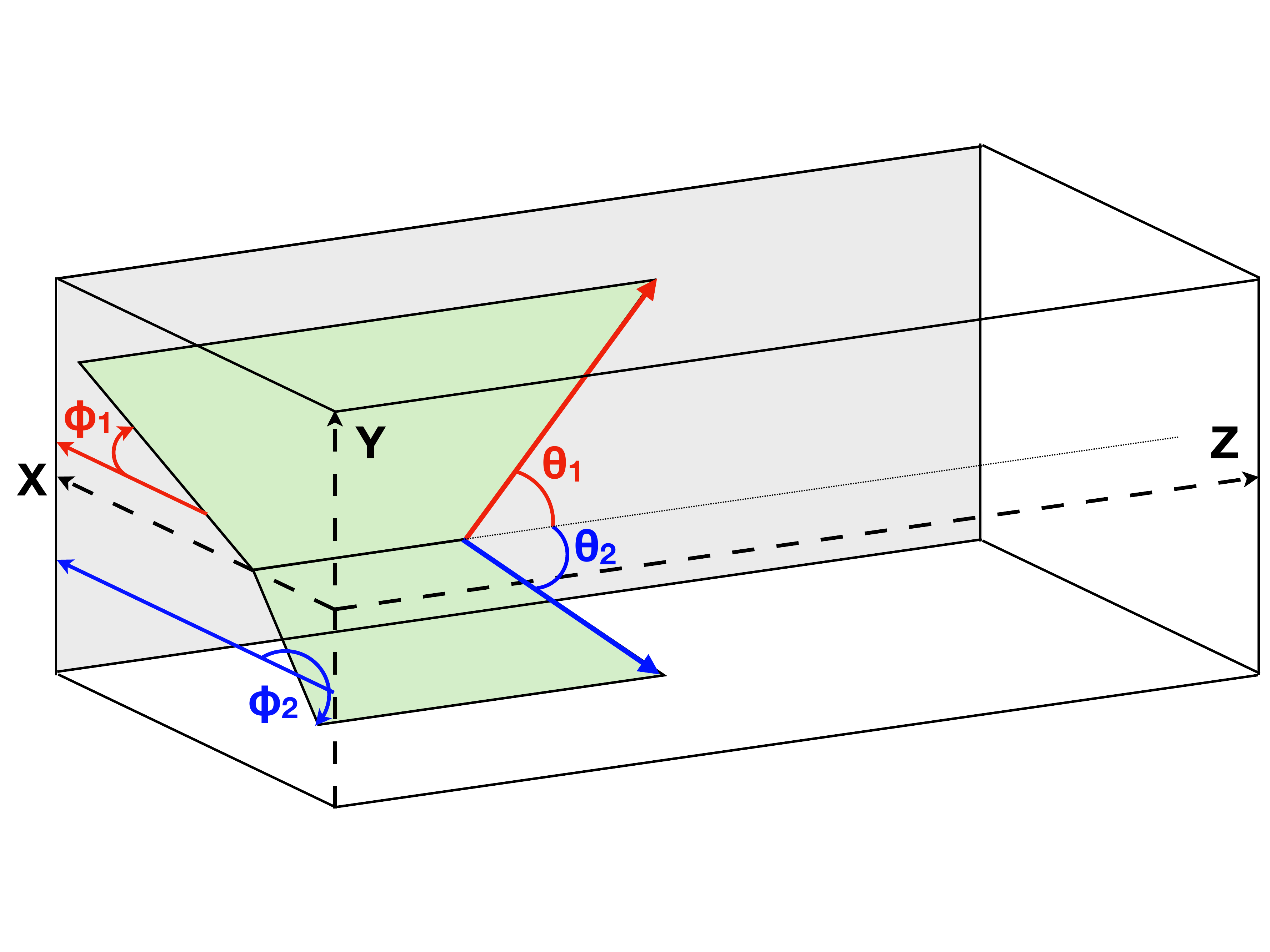} 
\caption{\label{angleDef} Description of $\phi$ and $\theta$ angles for each particle in MicroBooNE. $\phi$ is the angle of a track projected in the (X,Y) plane with respect to the X axis, and $\theta$ is the angle of a track with respect to the neutrio beam axis (Z axis).}
\end{figure}

For each individual track, the coordinates of the 3D points within \SI{15}{\centi \meter} of the vertex are averaged. The vector from the vertex to that mean point describes the path of the particle at short range. The angles $\phi$ , the projected angle in the (X,Y) plane, and $\theta$, the angle with respect to the beam axis, are evaluated for each track as shown in Fig. \ref{angleDef}.

\begin{figure}[!t]
\center
\subfigure[\label{ProtonPhiTracks}]{\begin{overpic}[width = 0.49\textwidth]{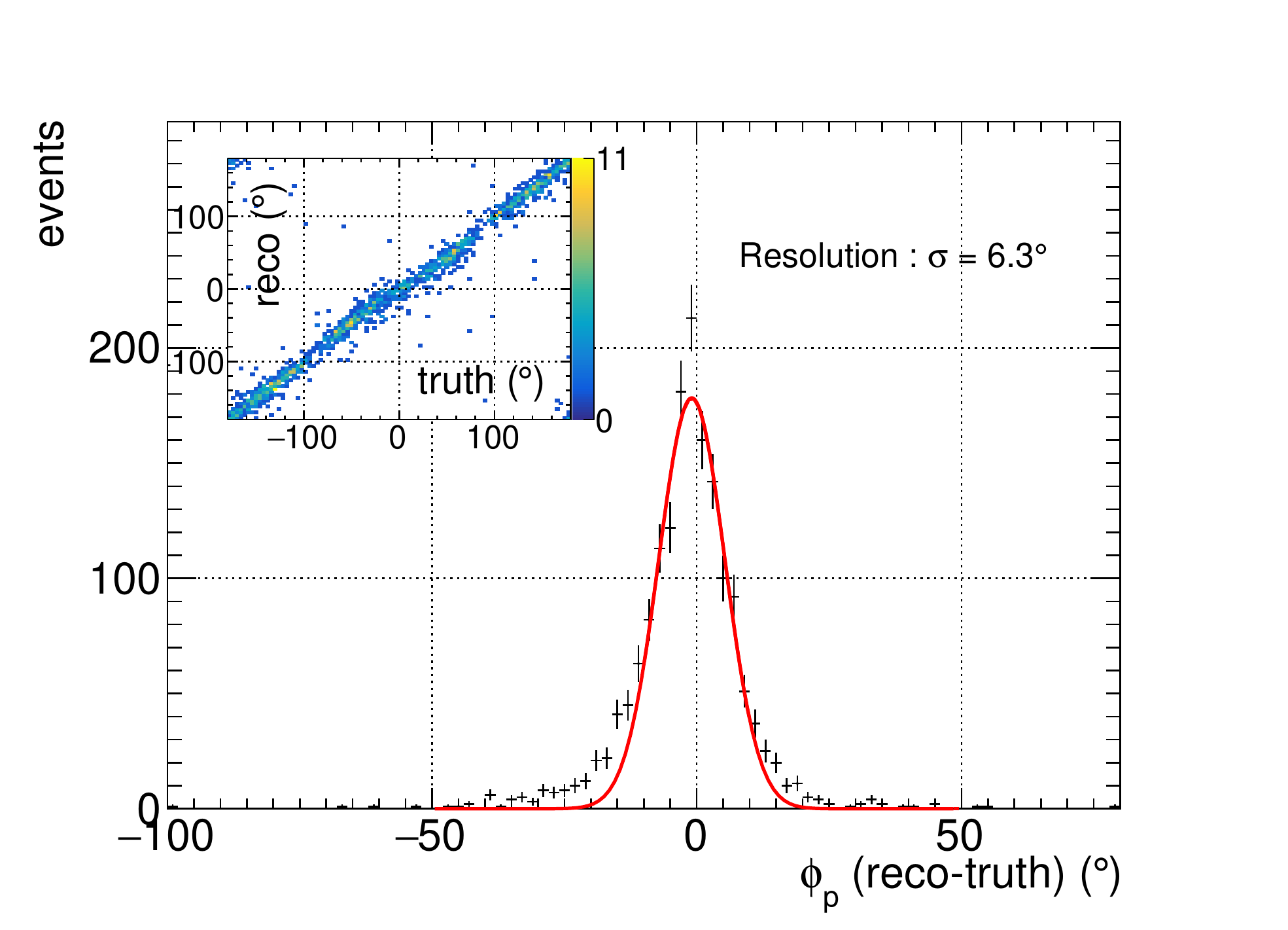}\put(13,65){\textbf{\scriptsize MicroBooNE Simulation}}\end{overpic} }
\subfigure[\label{MuonPhiTracks}]{\begin{overpic}[width = 0.49\textwidth]{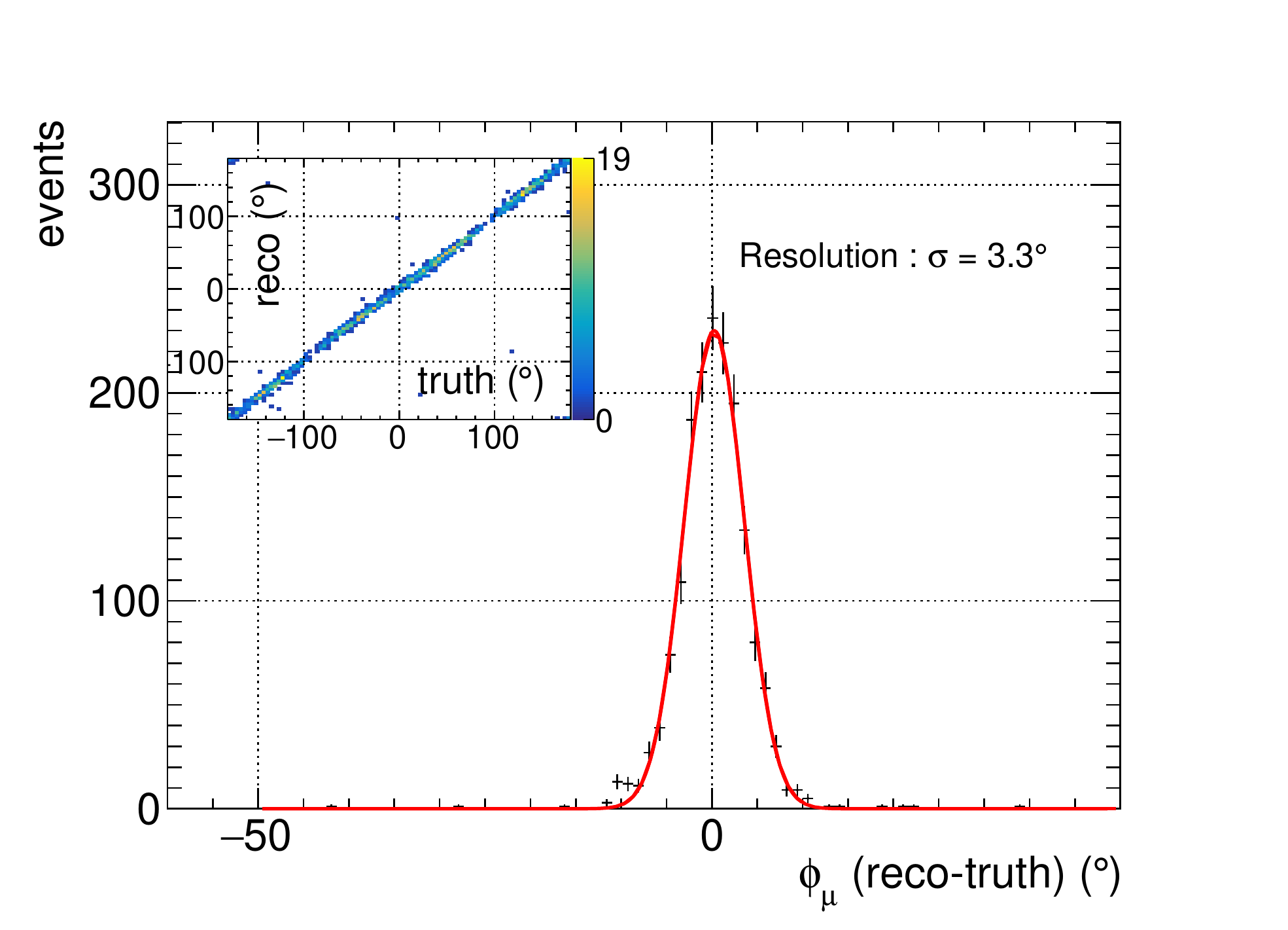}\put(13,65){\textbf{\scriptsize MicroBooNE Simulation}}\end{overpic} }
\caption{\label{PhiAnglesTracks} Error on the reconstruction of the $\phi$ angle for protons (a) and muons (b) from $1\mu 1p$ $\nu_\mu$ interactions. The red curve corresponds to a Gaussian fit of the central peak. A resolution of $6.3^{\circ}$ for the protons and $3.3^{\circ}$ for the muons is achieved. The inserts show the track by track correlation between the reconstructed and the true $\phi$ angles.}
\end{figure}

\begin{figure}[!t]
\center
\subfigure[\label{ProtonThetaTracks}]{\begin{overpic}[width = 0.49\textwidth]{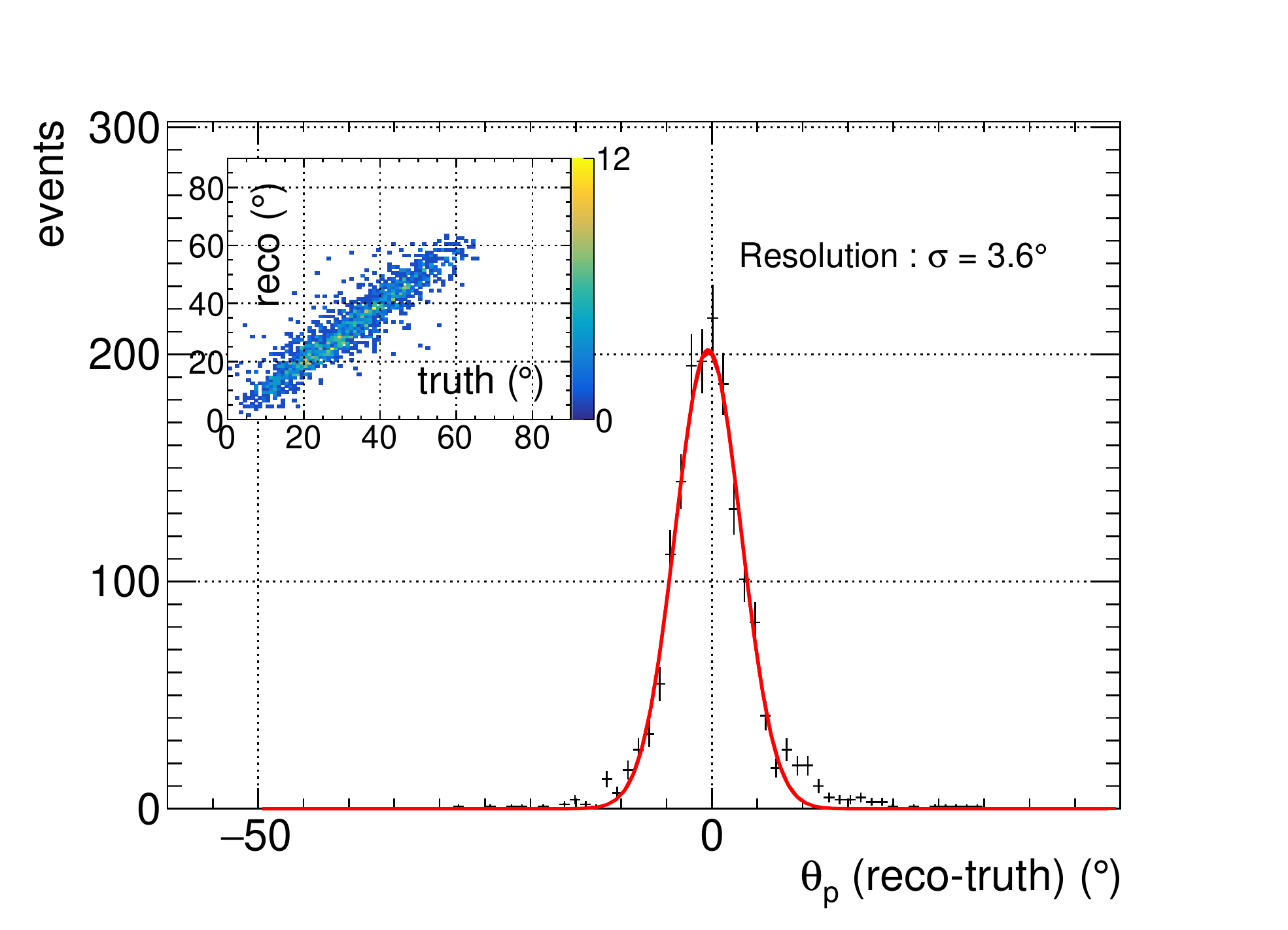}\put(13,65){\textbf{\scriptsize MicroBooNE Simulation}}\end{overpic} }
\subfigure[\label{MuonThetaTracks}]{\begin{overpic}[width = 0.49\textwidth]{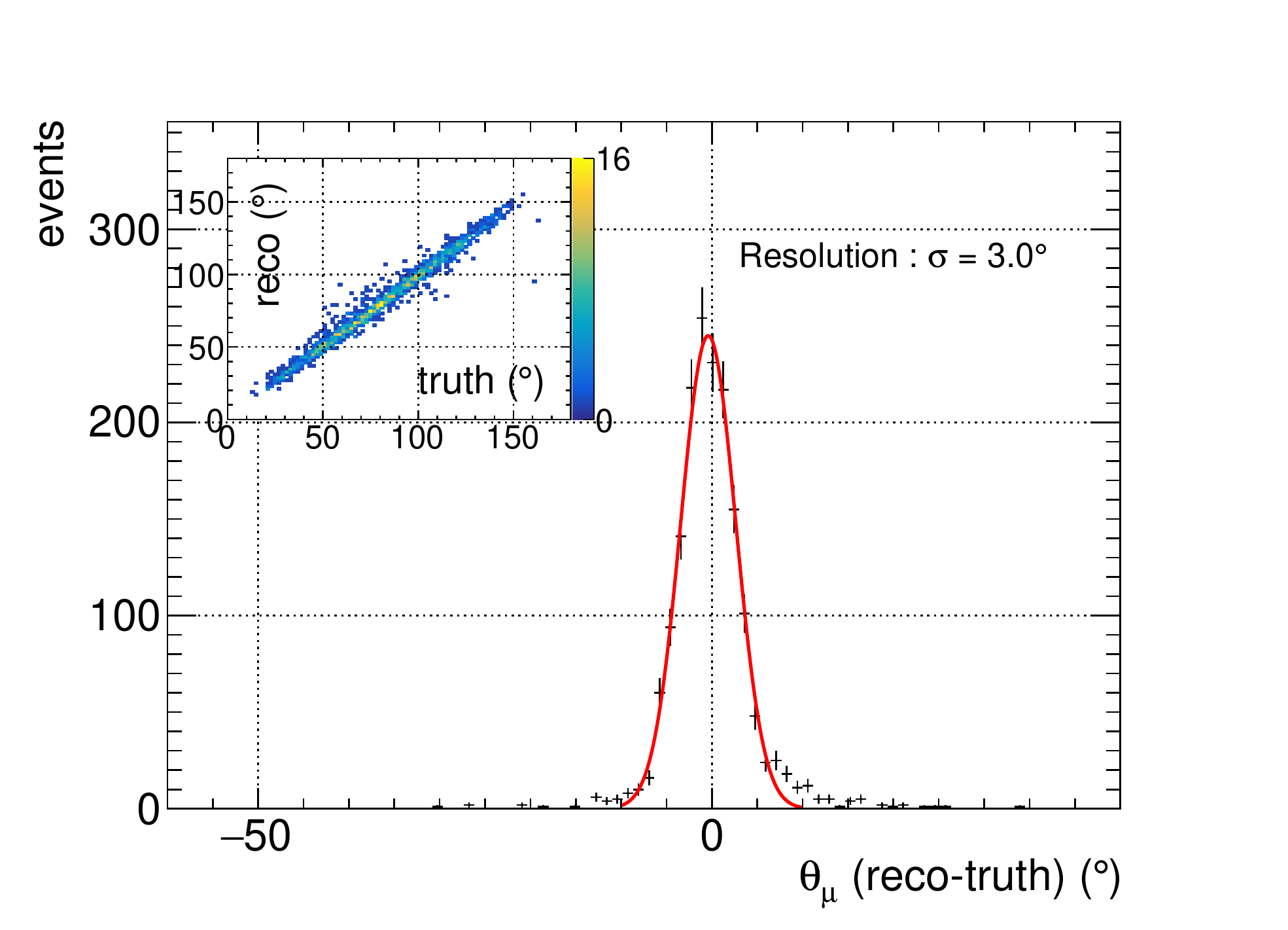}\put(13,65){\textbf{\scriptsize MicroBooNE Simulation}}\end{overpic} }
\caption{\label{ThetaAnglesTracks}Error on the reconstruction of the $\theta$ angle for protons (a) and muons (b) from $1\mu 1p$ $\nu_\mu$ interactions. The red curve corresponds to a Gaussian fit of the central peak. A resolution of $3.6^{\circ}$ for the protons and $3.0^{\circ}$ for the muons is achieved. The inserts show the track by track correlation between the reconstructed and the true $\theta$ angles.}
\end{figure}

Figures \ref{PhiAnglesTracks} and \ref{ThetaAnglesTracks} show the angular errors in the track reconstruction of protons and muons. 
Figures \ref{ProtonPhiTracks} and \ref{MuonPhiTracks} show the errors on the  $\phi$ angles for the protons and muons respectively. Resolutions in $\phi$ of $6.3^{\circ}$ for the protons and $3.3^{\circ}$ for the muons are achieved.

Figures \ref{ProtonThetaTracks} and \ref{MuonThetaTracks} show the errors on the  $\theta$ angles for the protons and muons respectively. Resolutions in $\theta$ of $3.6^{\circ}$ for the protons and $3.0^{\circ}$ for the muons are achieved.

\begin{figure}[!t]
\center
\begin{overpic}[width = 0.7\textwidth]{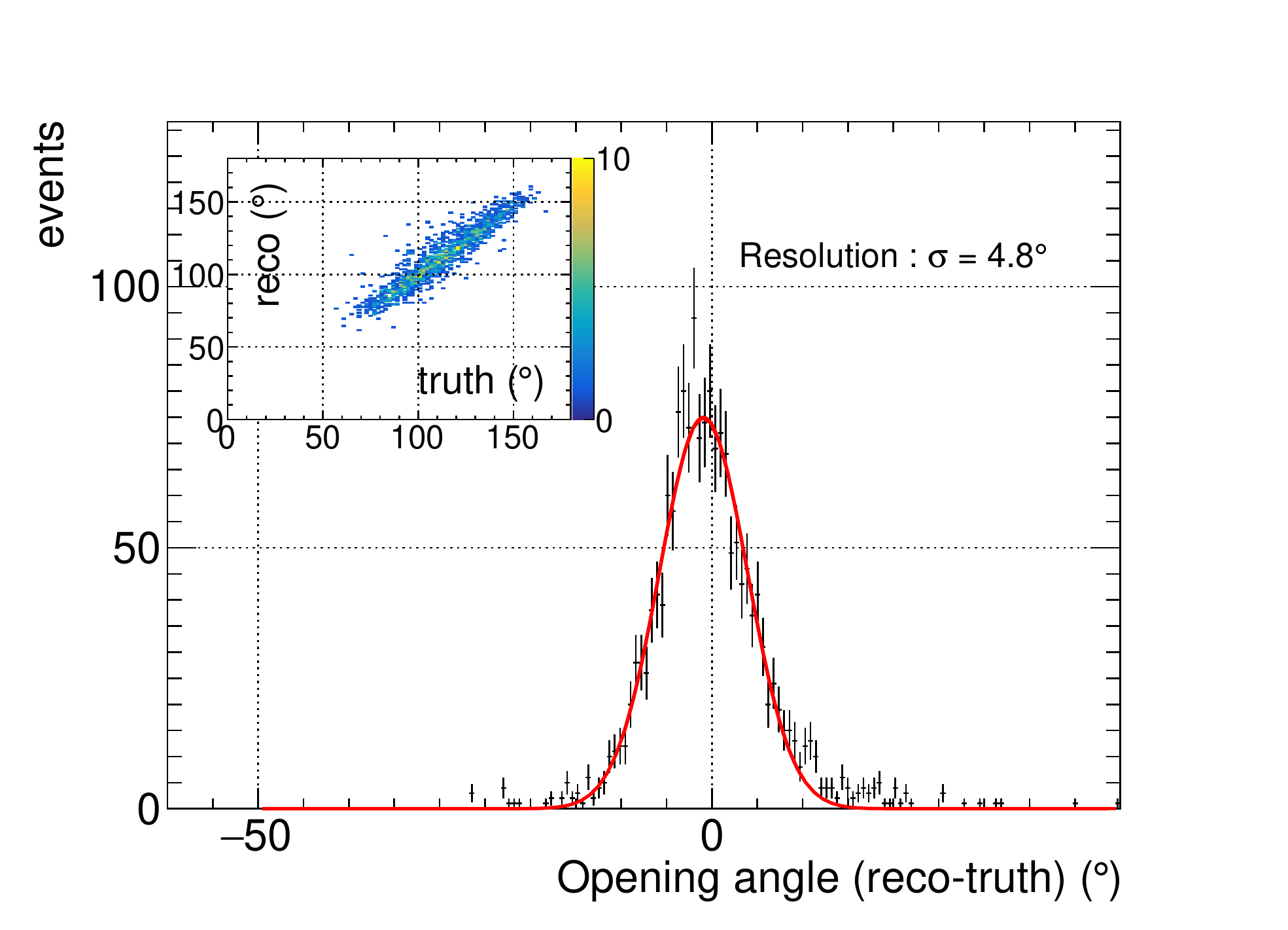}\put(13,65){\textbf{\scriptsize MicroBooNE Simulation}}\end{overpic}
\caption{\label{OpeningAngleTracks} Error on the reconstruction of the opening angle from $1\mu 1p$ $\nu_\mu$ interactions, defined as the angle between the two tracks. The red curve corresponds to a Gaussian fit of the central peak. A resolution of $4.8^{\circ}$ is obtained. The insert shows the event-by-event correlation between the reconstructed and the true opening angle.}
\end{figure}

Once the direction of each track has been computed, an opening angle can be evaluated by taking the angle between the direction of the two tracks.  Figure \ref{OpeningAngleTracks} shows the error made when reconstructing that opening angle. A fit to the central peak yields a resolution of $4.8^{\circ}$. The insert shows the linearity of the opening angle reconstruction.

\subsection{Energy Estimation \label{Ereco}}

The length of each track is the sum of two-point distances along the track. From the length of a track, a kinetic energy can be obtained, assuming a given particle identification, based on the stopping power of muons and protons in liquid argon \cite{NIST} \cite{PSTAR}.
As no particle identification has been performed yet, both energies are estimated for all tracks. It is left to the analyzers, downstream, to decide which one to use.

In order to estimate the neutrino energy reconstructed by this method, the attribution of proton or muon identification is performed as previously stated by using the average ionization and attributing the particle with the lower ionization as a muon and the one with the higher ionization as the proton. 

\begin{figure}[!t]
\center
\subfigure[\label{ProtonEres}]{\begin{overpic}[width = 0.49\textwidth]{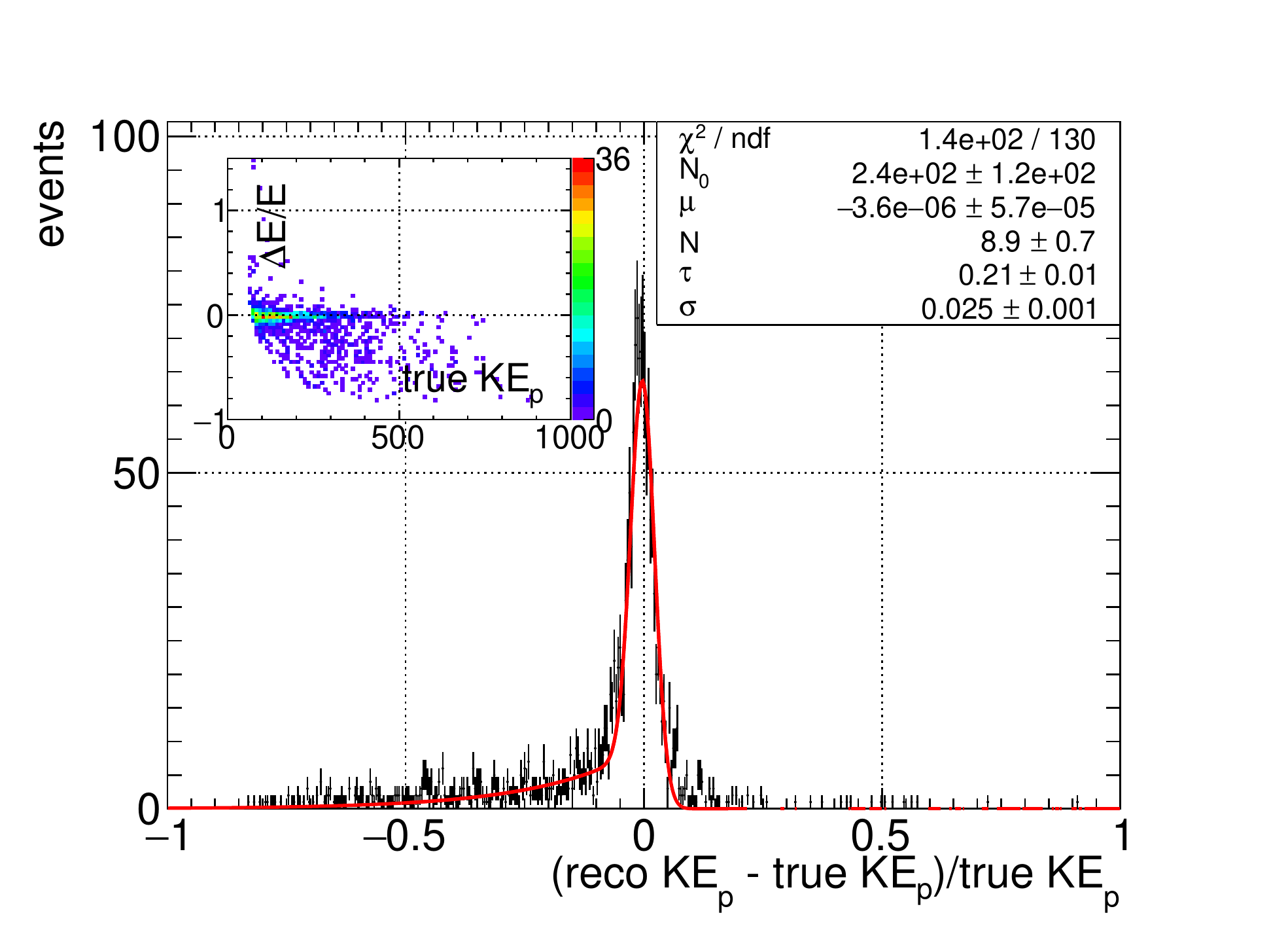}\put(13,65){\textbf{\scriptsize MicroBooNE Simulation}}\end{overpic}}
\subfigure[\label{MuonEres}]{\begin{overpic}[width = 0.49\textwidth]{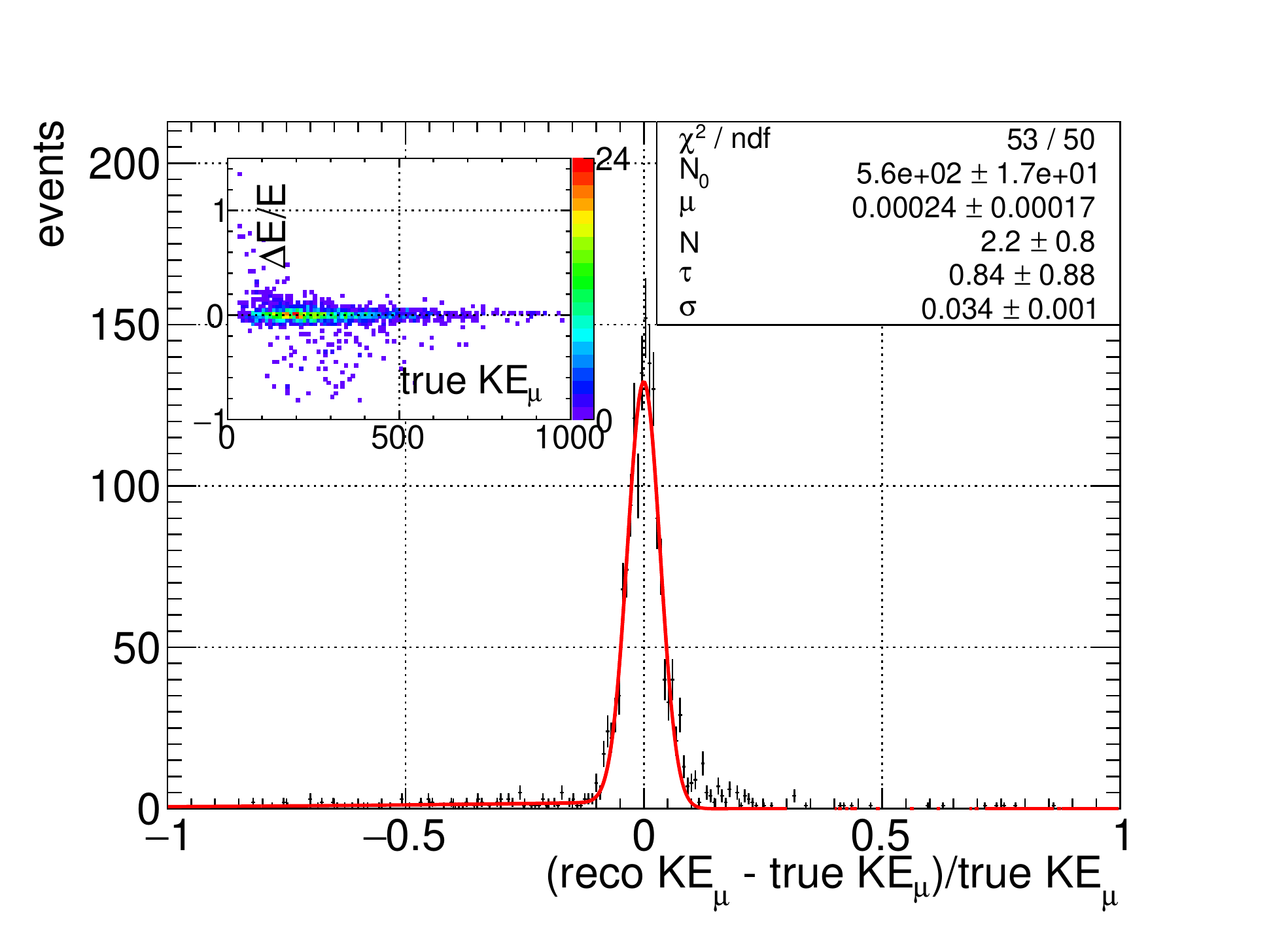}\put(13,65){\textbf{\scriptsize MicroBooNE Simulation}}\end{overpic}}
\caption{\label{singleParticleResolution} Relative difference between the reconstructed kinetic energy and the true kinetic energy as a function of the true kinetic energy at the single-track level for, respectively, proton (a) and muon (b) tracks from $1\mu 1p$ $\nu_\mu$ interactions.}
\end{figure}

Figure \ref{singleParticleResolution} shows the relative error made when reconstructing the energy of one of the two tracks.  
Figure \ref{ProtonEres} shows the proton resolution and \ref{MuonEres} shows the muon resolution. 

\begin{figure}[!t]
\center
\begin{overpic}[width = 0.7\textwidth]{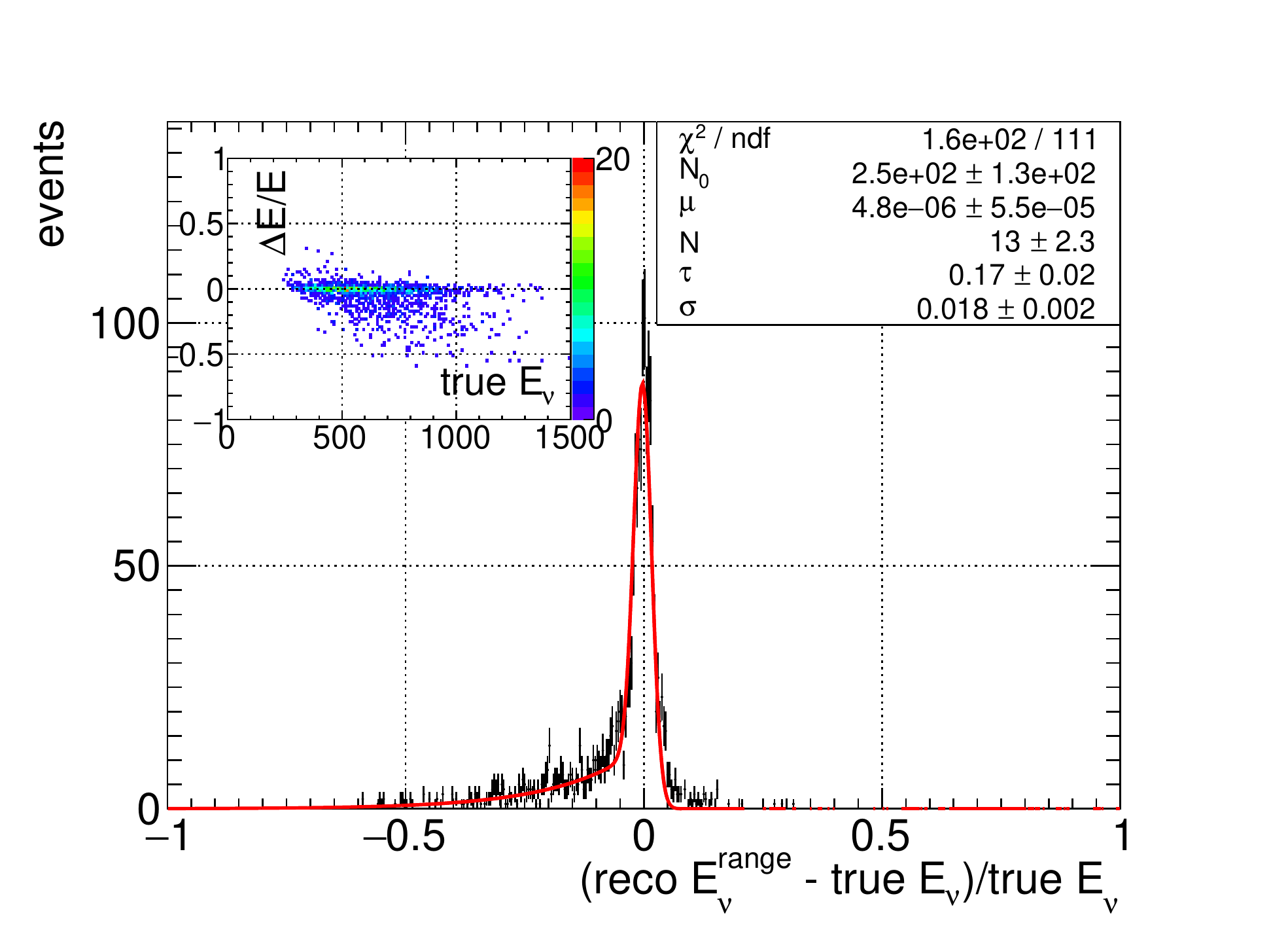}\put(13,65){\textbf{\scriptsize MicroBooNE Simulation}}\end{overpic}
\caption{\label{1DEnergyComparison} Relative difference between the neutrino energy reconstructed for the interaction and the true neutrino energy from $1\mu 1p$ $\nu_\mu$ interactions. The central peak shows a resolution of $(1.8 \pm 0.2) \% (statistical errors only)$.}
\end{figure}

The distributions show a relative error centered on $(KE^{\mathrm{reco}}-KE^{\mathrm{true}})/KE^{\mathrm{true}} = 0$, with a tail towards lower values corresponding to events with missing reconstructed energy. The 1-D distributions in figures \ref{ProtonEres}, \ref{MuonEres} and \ref{1DEnergyComparison} are fitted by the following function :
\begin{align}
f(x) &= \left(N_{0}\times \delta(x-\mu) + N \times H(\mu-x) \times e^{(x-\mu)/\tau}\right) * \dfrac{1}{\sigma\sqrt{2\pi}}e^{-\dfrac{x^2}{2\sigma^{2}}} \label{fitFunction}
\end{align}
Where $x = (KE^{\mathrm{reco}}-KE^{\mathrm{true}})/KE^{\mathrm{true}}$ is the relative error on the reconstructed events, and $H(\mu-x)$ is the Heaviside step function, which takes values of $1$ for $x < \mu$, and $0$ for $x \ge \mu$. In a case of a perfect reconstruction, all events would follow a Dirac distribution centered on 0. In practice, some particles travel less than their original kinetic energy would suggest under a pure Bethe-Bloch behaviour, because of non-ionizing energy loss for instance, this behaviour leads to a tail to negative values beside the Dirac distribution, which we model as an exponential of characteristic parameter $\tau$. The relative normalization of the tail to the Dirac distribution is given by the parameters $N_{0}$ and $N$. The sum of these two distributions is then convoluted with a unitary Gaussian of width $\sigma$ that represents the resolution of the reconstruction. We also account for a systematic loss in energy reconstruction by shifting the whole distribution by $\mu$.

The central peak of the proton distribution has a resolutions of $(2.5 \pm 0.1)\%$.
The distribution of the proton energy error has a significant tail where the reconstructed energy is lower than the true kinetic energy; this can be due to misreconstructions, and to physical processes where the reconstructed length is not representative of the true kinetic energy, such as energy loss due to nuclear effects or a hard scatter of the proton along its path. 
 The distribution of the muon energy error, on the other hand, does not present such a tail at lower energies. The smaller tail to higher energies  corresponds to a fraction of the Michel electron being added to the muon track.\\
The central peak of the muon distribution shows a resolutions of $(3.4 \pm 0.1) \%$
The inserts show the relative errors as a function of the kinetic energies. The distributions are mainly flat in kinetic energies, showing that the resolution remains similar whatever the energy of the particle.\\
 
Figure \ref{1DEnergyComparison} shows the relative error made in reconstructing the full energy of the neutrino. We use here the range-based neutrino energy definition :
\begin{align} 
E_{\nu}^{range} &= KE_{p}+KE_{\mu}+m_{\mu}+m_{p}+B-m_{n},
\end{align}
where $B \sim \SI{40}{\mega \electronvolt}$ is an effective nuclear binding energy \cite{PhysRevC.74.054316}. We compare the ranged-based energy to the true neutrino energy to obtain the fractional reconstruction resolution.
The distribution is fitted with the function described in equation \ref{fitFunction}. The peak of the distribution shows a resolution of $(1.8 \pm 0.2)\%$. The events in the  tail account for $40\%$ of the events, and are driven by the lost energy when reconstructing proton tracks, on one part, and of non-ionizing energy loss processes such as neutron emission.

\begin{figure}[!t]
\center
\subfigure[\label{linearityTest}]{\begin{overpic}[width = 0.7\textwidth]{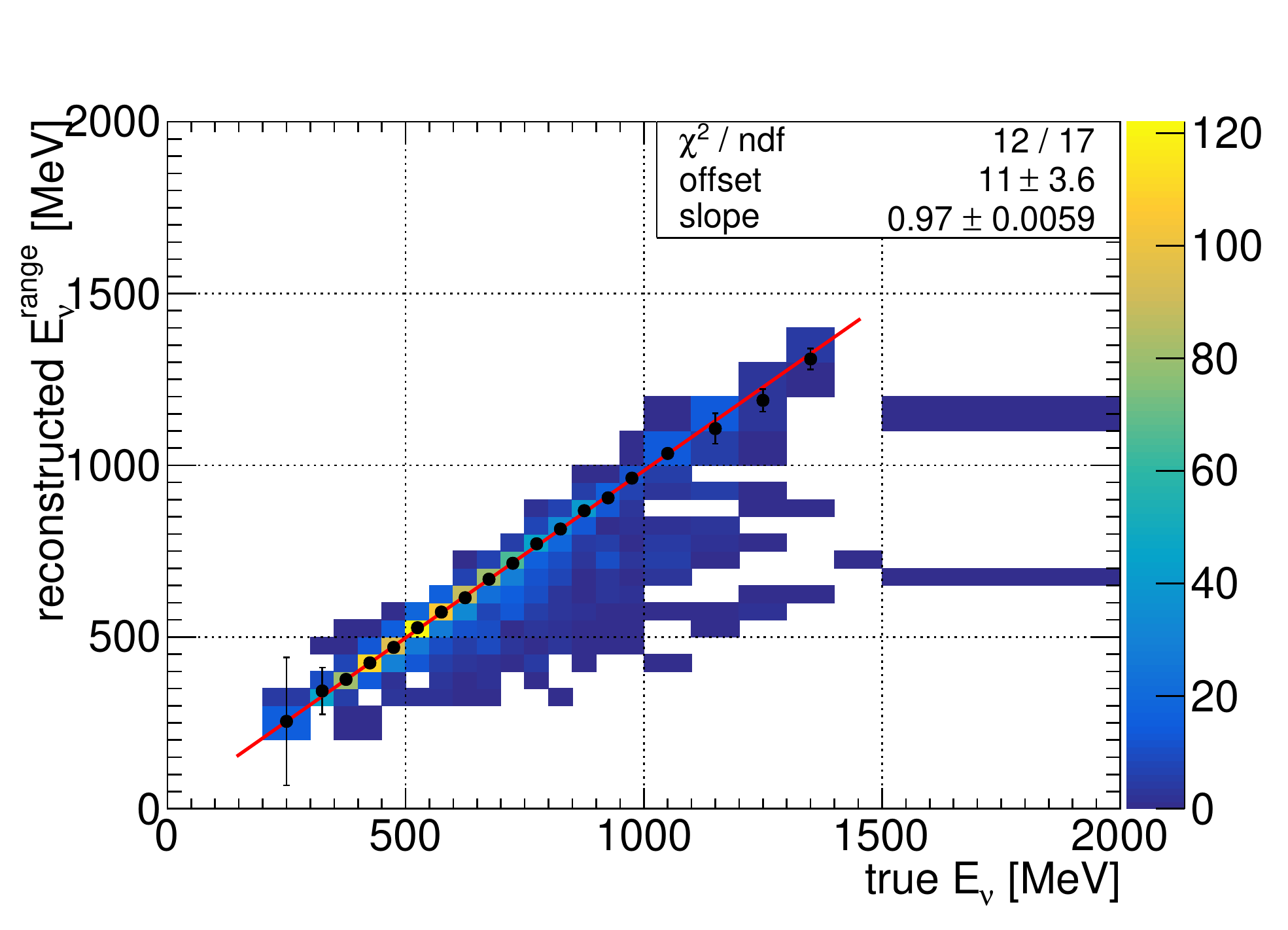}\put(13,65){\textbf{\scriptsize MicroBooNE Simulation}}\end{overpic}}
\subfigure[\label{EresVSEgraph}]{\begin{overpic}[width = 0.7\textwidth]{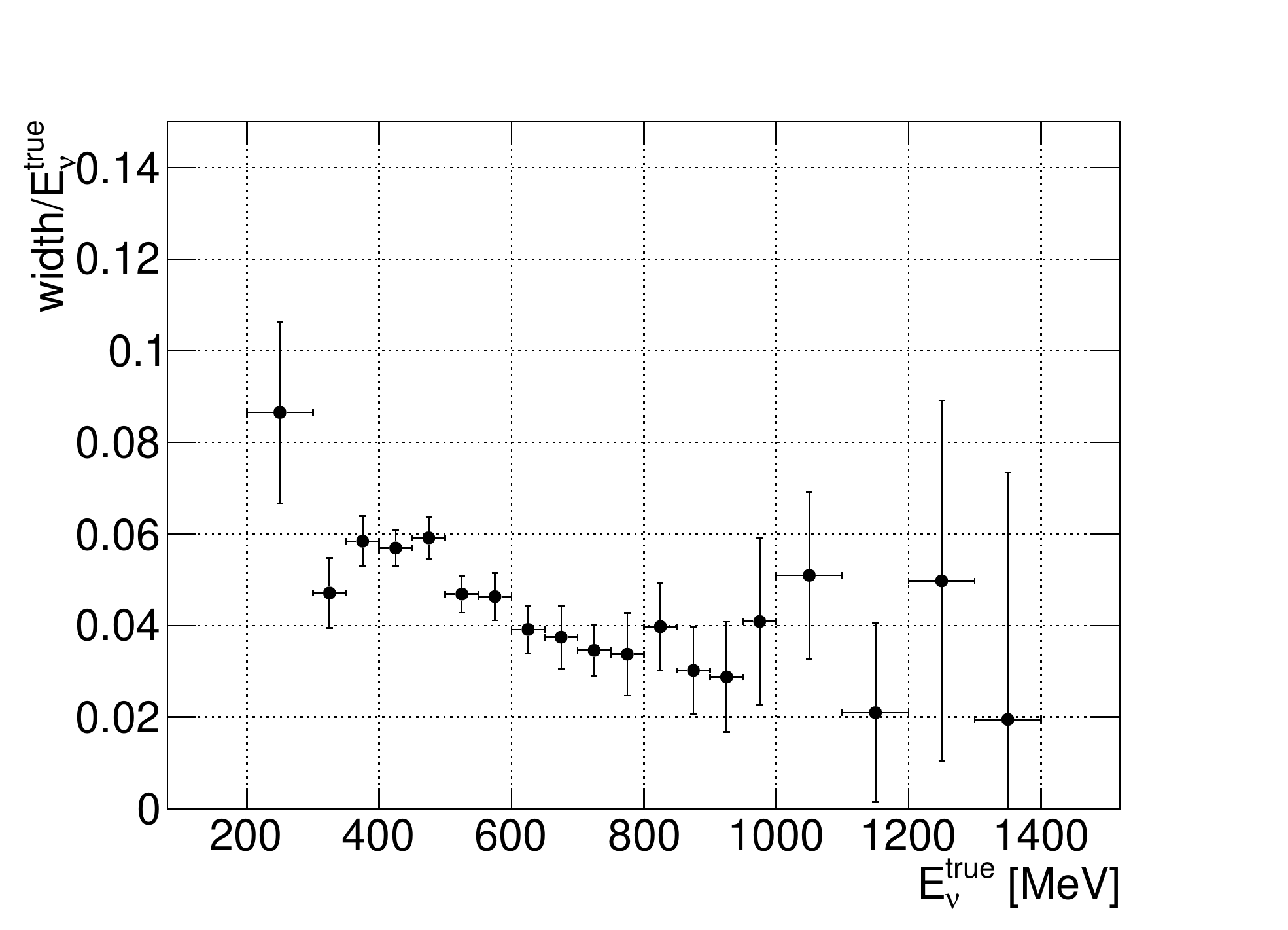}\put(13,65){\textbf{\scriptsize MicroBooNE Simulation}}\end{overpic}}
\caption{\label{linearityEstimation} (a) : Comparison of the reconstructed neutrino energy to the true MC neutrino energy for $1\mu 1p$ $\nu_\mu$ interactions. (b) Evolution of the resolution as a function of the true neutrino energy.}
\end{figure}

Figure \ref{linearityTest} shows a comparison of the true and reconstructed neutrino energy. Each slice in true energy is fitted by a Gaussian around its mean value using the Minuit2 minimization package \cite{Minuit2}. The fit results are shown as the black dots, and the errors on these dots correspond to the errors on the Gaussian's mean. 
A linear fit performed on the results shows a strong linearity, with a slope factor of $0.97 \pm 0.006$ and an offset of $(11 \pm 3)\, \si{\mega \electronvolt}$, for a neutrino energy range of $[200-1400] \si{\mega \electronvolt}$.\\

Figure \ref{EresVSEgraph} shows the evolution of the fractional resolution ($\sigma$/true neutrino energy) as a function of true neutrino energy. The errors are the errors on the $\sigma$ parameters estimated by the Gaussian fit divided by the true neutrino energy. The relative resolution shows an upward trend towards low energies.

\subsection{Efficiency Estimation \label{EffReco}}

\begin{figure}[!t]
\center
\subfigure[\label{EfficiencPlots}]{\begin{overpic}[width = 0.49\textwidth]{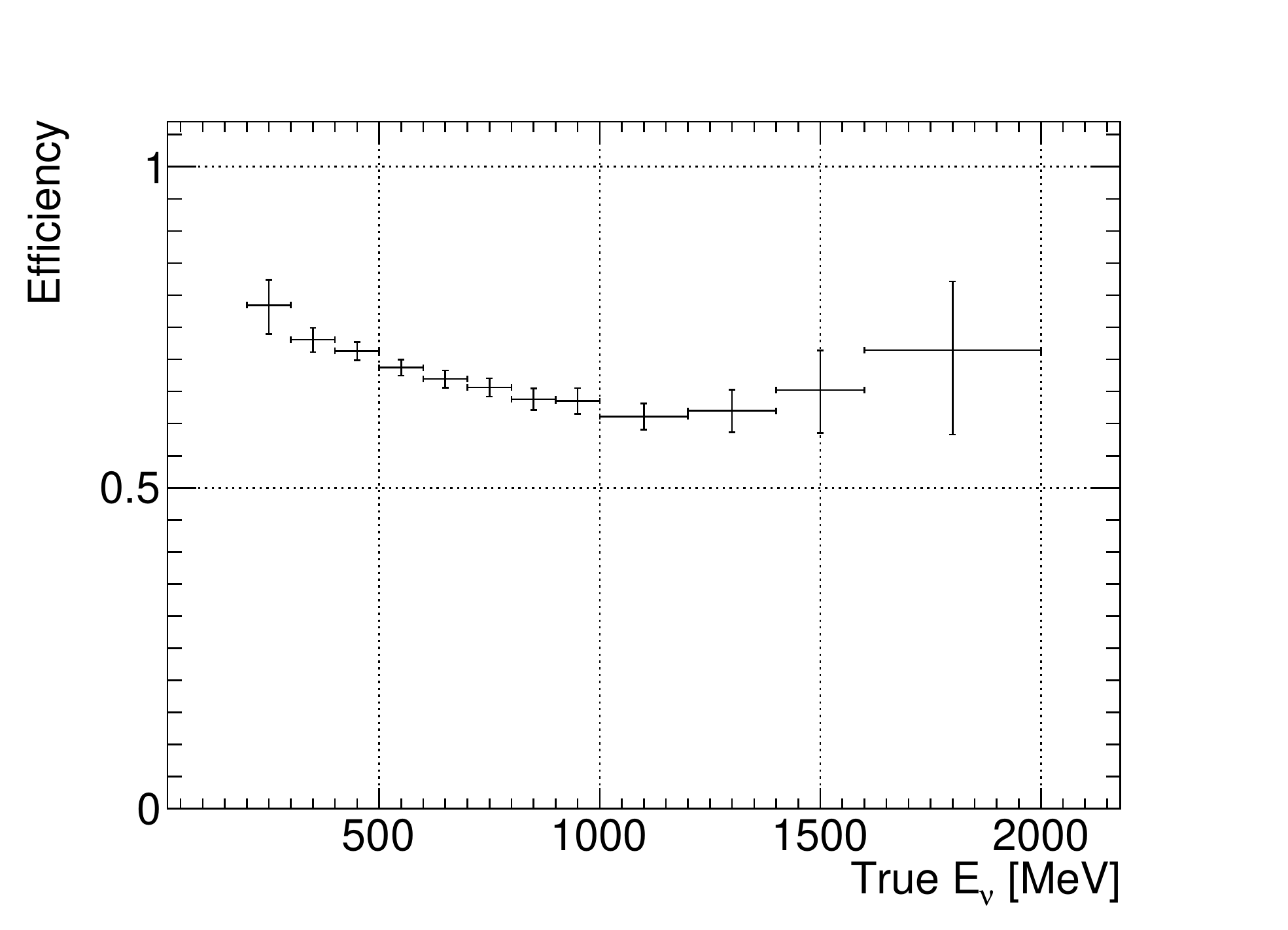}\put(13,65){\textbf{\scriptsize MicroBooNE Simulation}}\end{overpic}}
\subfigure[\label{EfficiencOpenAngle}]{\begin{overpic}[width = 0.49\textwidth]{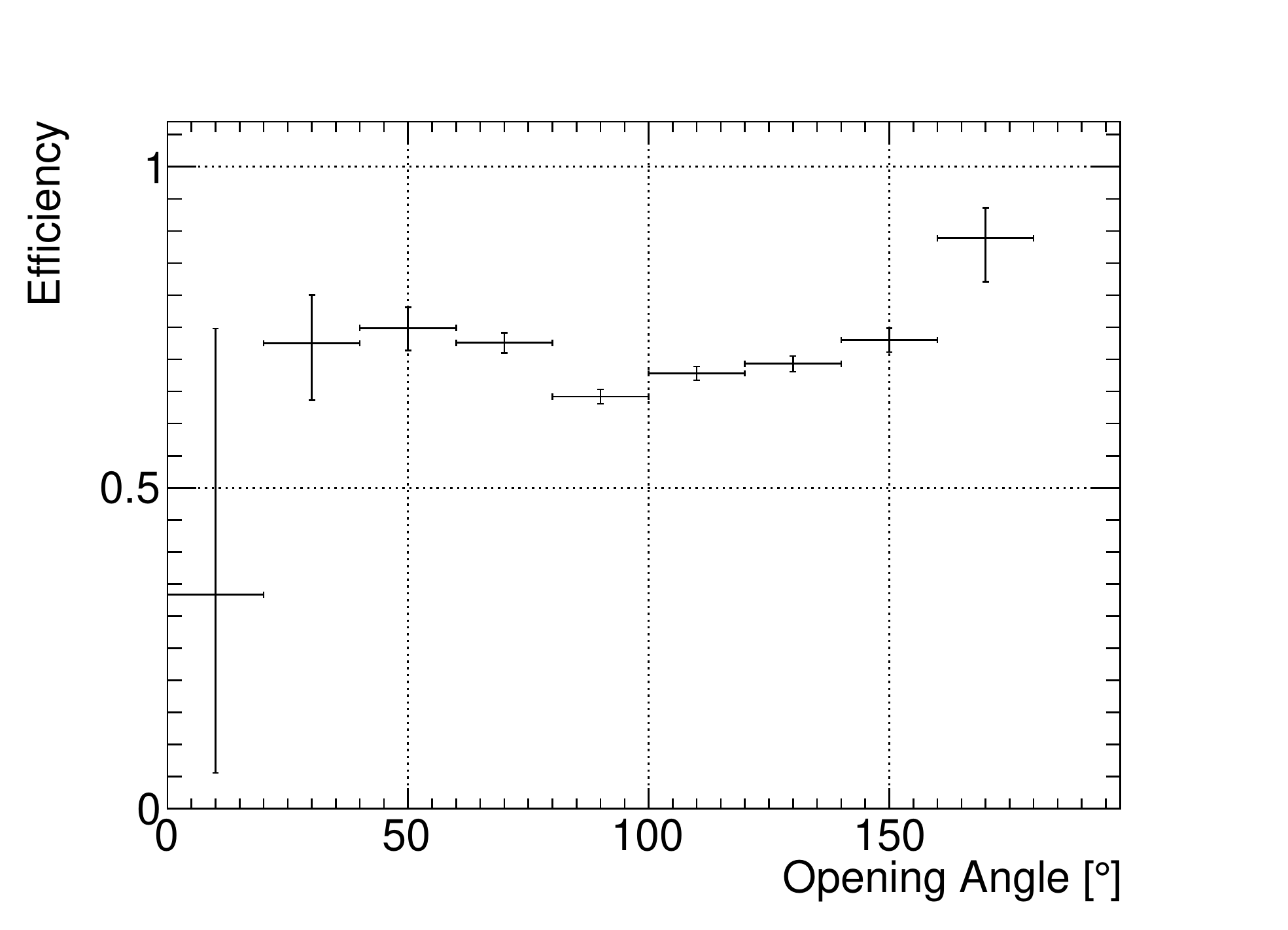}\put(13,65){\textbf{\scriptsize MicroBooNE Simulation}}\end{overpic}}
\caption{ Tracking efficiency as a function of the true neutrino energy for events that are well reconstructed, i.e. for which the reconstruction reaches the end of the tracks and with only two found tracks above \SI{5}{\centi \meter} from $1\mu 1p$ $\nu_\mu$ interactions.}
\end{figure}

Figure \ref{EfficiencPlots} shows the evolution of the tracking efficiency with respect to the true neutrino energy. For each energy bin, the efficiency is calculated as the ratio of the well reconstructed $1\mu 1p$ $\nu_\mu$ interactions fully contained in the fiducial volume of the detector that have their vertex well reconstructed (within \SI{3}{\cm} of the true vertex) divided by $1\mu 1p$ $\nu_\mu$ interactions fully contained in the fiducial volume of the detector that have their vertex well reconstructed. The criterion for good reconstruction is set using the diagnostic tool described earlier, and not a fraction of reconstructed length based on truth-level information. This choice allows us to use the same tools that will be used in the actual data analysis for estimating the realistic performance of the track reconstruction. The average efficiency of the track reconstruction is $69\%$ over the full energy range; however, a downward trend is visible,  which is linked to the increased probability of encountering a failure as the track length increases, such as unresponsive wires or wire signal deconvolution issues. \\

Figure \ref{EfficiencOpenAngle} shows the same efficiency, as a function of the true opening angle between the proton and the muon. The efficiency is mostly flat, with a slight upward trend towards the higher opening angle values, where the tracks are less likely to overlap in one of the three TPC views, thus making their reconstruction more efficient.

\begin{figure*}[!t]
\center

\subfigure[\label{ADCimage_simple}]{\begin{overpic}[width = 0.46\textwidth]{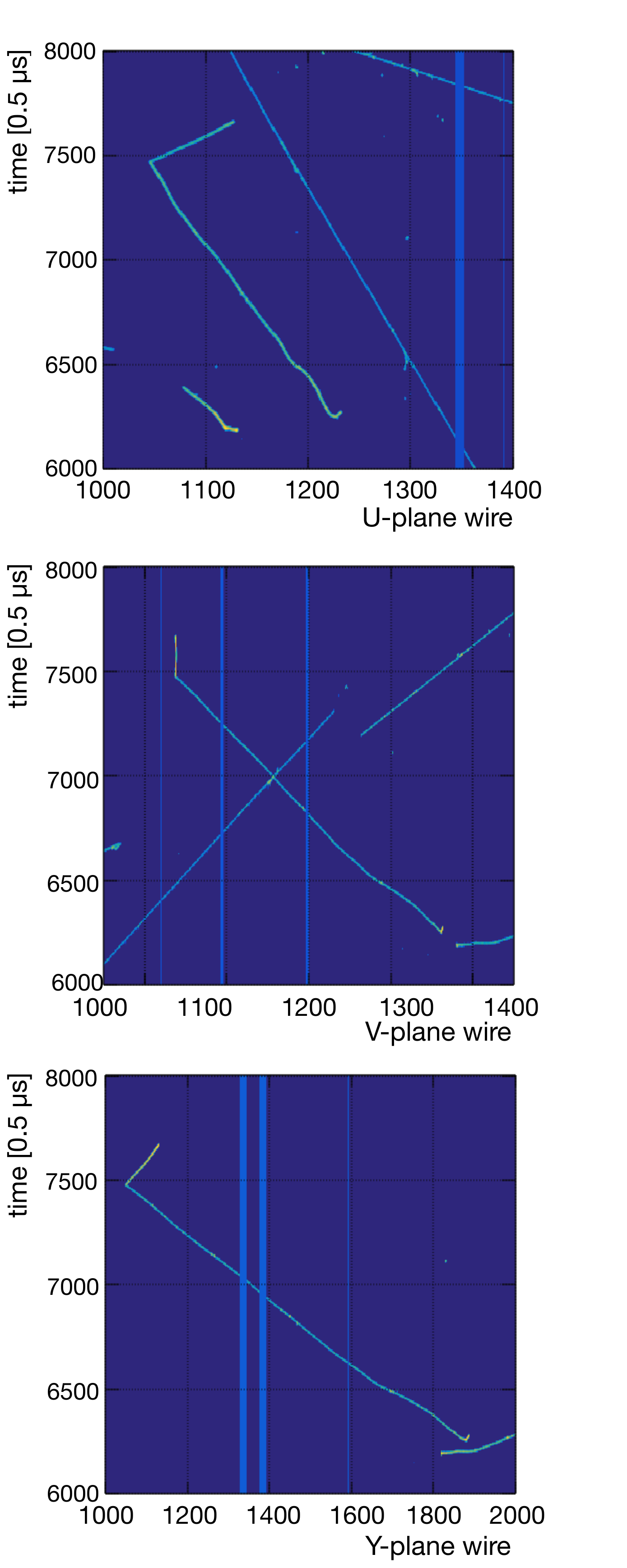}\put(7,97.5){\textbf{\scriptsize MicroBooNE Simulation}}\end{overpic}}
\subfigure[\label{ADCimage_reco}]{\begin{overpic}[width = 0.46\textwidth]{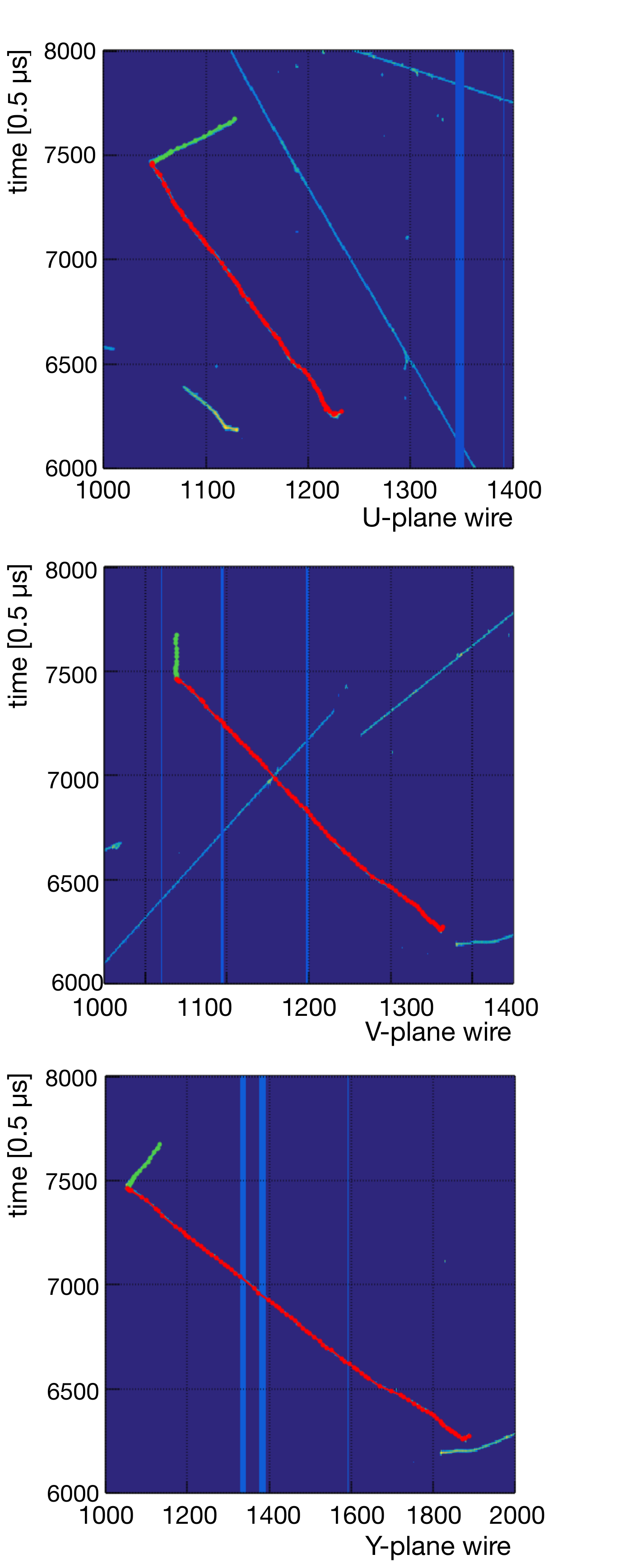}\put(7,97.5){\textbf{\scriptsize MicroBooNE Simulation}}\end{overpic}}

\caption{\label{recoedMCvertex} Example of a reconstructed MC event : 
(a) ADC image of a \SI{975}{\mega \electronvolt} simulated $1\mu 1p$ neutrino event producing a \SI{603}{\mega \electronvolt} muon and a \SI{226}{\mega \electronvolt} proton. 
(b) Reconstructed tracks are overlaid on top of the ADC image. The event is reconstructed as a \SI{627}{\mega \electronvolt} muon (red) and  a \SI{221}{\mega \electronvolt} proton (green).}
\end{figure*}

As a reconstruction example, the event shown in Fig. \ref{recoedMCvertex} is a \SI{975}{\mega \electronvolt} neutrino (range-based energy), producing a \SI{603}{\mega \electronvolt} muon and a \SI{226}{\mega \electronvolt} proton. 
Figure \ref{ADCimage_simple} shows the three ADC images corresponding to the view of each plane cropped around the neutrino interaction. 
Figure \ref{ADCimage_reco} shows the projections  of reconstructed tracks for each plane overlaid on top of the corresponding ADC images. 
The straight vertical light blue lines correspond to the unresponsive wires. 
The reconstructed energies are respectively \SI{627}{\mega \electronvolt} for the muon track (red dots) and \SI{221}{\mega \electronvolt} for the proton track (green dots). 
The reconstructed range-based energy is \SI{993}{\mega \electronvolt} and differs by $+1.8\%$ from the true range-based energy.

\clearpage
\section{Comparison to the Pandora-based reconstruction}

\begin{figure}[!t]
\center
\begin{overpic}[width = 0.7\textwidth]{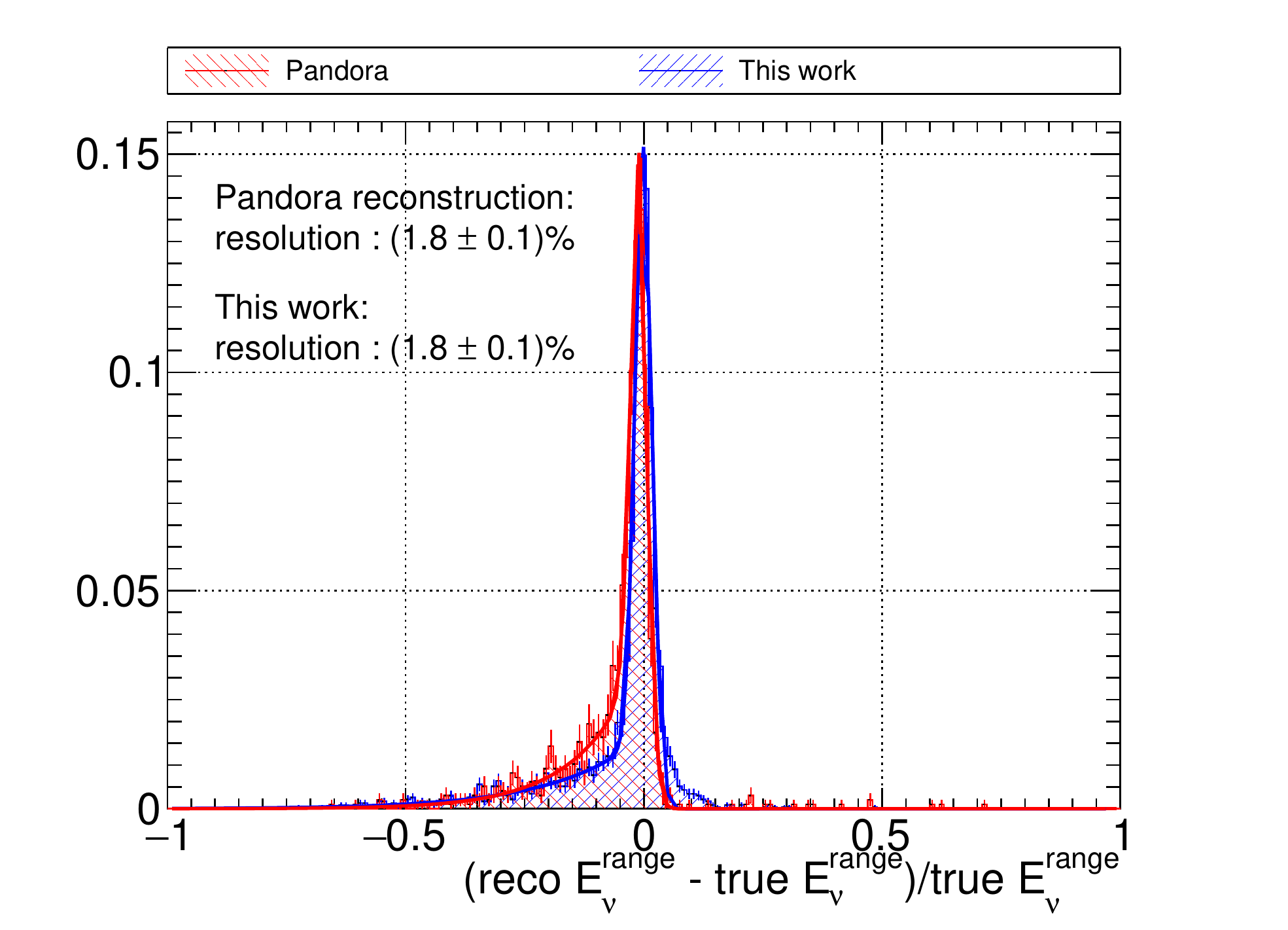}\put(13,70){\textbf{\scriptsize MicroBooNE Simulation}}\end{overpic}
\caption{\label{CompPandoraRes} Comparison of the resolutions of reconstructed CCQE $\nu_\mu$ interactions with a $1\mu1p$ topologies using either the Pandora package (red) or this reconstruction (blue). Both distributions are unit-normalized. A fit to the central peak shows similar resolutions.}
\end{figure}

The reconstruction presented here makes use of the detector readout as an image and as such uses image processing tools such as convolutional neural networks and image processing. Other reconstruction algorithms are used in MicroBooNE, including the Pandora pattern recognition package \cite{PandoraPaper}. 
Figure \ref{CompPandoraRes} shows the distribution of the fractional residuals of the reconstructed neutrino energy for our target interaction topology $\nu_\mu$ CCQE $1\mu1p$ topology both for the Pandora reconstruction package and the reconstruction algorithm described in this paper.
The two distributions are unit-normalized; a fit to the central peak region shows that the two methods have the same core resolution within the Monte Carlo simulation statistical errors, of the order of $(1.8 \pm 0.1)\%$. The reconstruction described here offers a complementary and independent alternative to the Pandora reconstruction package currently in use in LArTPC experiments, and provides similar reconstruction performance for two-track events.


\section{Conclusion}

We present a code package for three-dimensional event reconstruction of two-track events in LArTPCs.  The code is publicly available on \mbox{GITHUB \cite{githubcode}.} This reconstruction uses computer vision and clustering tools to find 3D-consistent vertices, and a 3D stochastic best-neighbor search to reconstruct tracks emerging from these vertices.  While we have discussed the algorithms within the context of event reconstruction in the MicroBooNE detector, this package can be ported to other LArTPC detectors with only minor changes, such as the geometry.  Because future experiments of the Fermilab SBN program have similar LArTPC designs and run in the same BNB neutrino beam-line, the code is easily adaptable for SBN use.   DUNE analyses involving beam-induced low energy neutrinos as well as atmospheric neutrinos will also find aspects of the code to be applicable.\\

The main parameters that affect the performance of the vertexing algorithm are the outgoing proton and muon energies and their opening angle. 
The vertexing algorithm is also affected by the performance of upstream reconstruction stages such as the SSNet labelling, the cROI finding, and the cosmic ray pixel tagger.
From the output of the vertexing stage, the track finding algorithm is mainly affected by the unresponsive detector regions. The dependence of the efficiency on the energy is simply due to an increased probability of crossing such a region as the track length increases.

The performance of the vertex reconstruction was optimized to precisely place vertices on 3D-consistent kinks and track-shower boundaries with a maximum efficiency. The track reconstruction is then run for all  found vertices. The performances of the track reconstruction are optimized with the objective of finding nearby tracks with small opening angles, and to maximize the length of the reconstructed tracks by following a 3D-consistent path of non-zero charge deposition. A smoothing algorithm is then applied to the found set of points and finally, a self-diagnostic is performed to ensure that the reconstructed variables are relevant, and rejects the vertices for which the reconstruction failed to follow the track to its end to increase the purity of the final sample of events.  The efficiencies  for neutrino events in the $200-1250$\si{\mega \electronvolt} of the vertex finding is $56\%$ and that of the track reconstruction is $69\%$. The spatial resolution of the vertex finding algorithm is of the order of the wire spacing of MicroBooNE, and the neutrino energy resolution of 2-track truth-level CCQE events reconstructed using this tracking method is $(1.8 \pm 0.2) \%$, in the $200-1000$ MeV energy range.

Further work will include allowing the tracker to recover tracks by improving its ability to cross through unresponsive regions. The local ionization will also be exploited to better pinpoint the end of muon tracks and separate the Michel electron. A spatial correction of the reconstructed points based on a precise measurement of the space-charge effect inducing inhomogeneities of drift field  and distorting the reconstructed tracks will also be performed, further improving the performance of this reconstruction. Furthermore, the systematic impact of the uncertainties from event generator, as well as from the detector simulation, still need to be investigated. These are mostly analysis dependent and will be described in the communications of the analyses making use of the reconstruction described above. Systematic effects from the event generator can be reweighed to account for various models describing the neutrino interactions and the propagation of the daughter particles, while detector effect will require the use of simulated events for many universes describing the range of possible detector configuration. As these studies are beyond the scope of this work, they are not presented here.

\section*{Acknowledgements}

This material is based upon work supported by the following: the U.S. Department of Energy, Office of Science, Offices of High Energy Physics and Nuclear Physics; the U.S. National Science Foundation; the Swiss National Science Foundation; the Science and Technology Facilities Council of the United Kingdom; and The Royal Society (United Kingdom). Additional support for the laser calibration system and cosmic ray tagger was provided by the Albert Einstein Center for Fundamental Physics. Fermilab is operated by Fermi Research Alliance, LLC under Contract No. DE-AC02-07CH11359 with the United States Department of Energy.

\end{document}